\documentclass[traditabstract]{aa} 
\usepackage{graphicx}
\usepackage{txfonts}
\begin{document}
\title{Radio-to-$\gamma$-ray Monitoring of the Narrow-Line Seyfert 1 Galaxy PMN~J0948$+$0022 from 2008 to 2011}

\author{L. Foschini\inst{1}, 
E. Angelakis\inst{2}, 
L. Fuhrmann\inst{2}, 
G. Ghisellini\inst{1}, 
T. Hovatta\inst{3}, 
A. Lahteenmaki\inst{4}, 
M. L. Lister\inst{5}, 
V. Braito\inst{1}, 
L. Gallo\inst{6}, 
T. S. Hamilton\inst{7}, 
M. Kino\inst{8}, 
S. Komossa\inst{2}, 
A. B. Pushkarev\inst{9,10,2}, 
D. J. Thompson\inst{11}, 
O. Tibolla\inst{12}, 
A. Tramacere\inst{13}, 
A. Carrami\~nana\inst{14},
L. Carrasco\inst{14}, 
A. Falcone\inst{15}, 
M. Giroletti\inst{16}, 
D. Grupe\inst{15}, 
Y.~Y. Kovalev\inst{17,2}
T. P. Krichbaum\inst{2}, 
W. Max-Moerbeck\inst{3},
I. Nestoras\inst{2}, 
T. J. Pearson\inst{3},
A. Porras\inst{14},
A. C. S. Readhead\inst{3},
E. Recillas\inst{14},
J. L. Richards\inst{5},
D. Riquelme\inst{18},
A. Sievers\inst{18},
J. Tammi\inst{4}, 
M. Tornikoski\inst{4}, 
H. Ungerechts\inst{18},
J. A. Zensus\inst{2},
A. Celotti\inst{19,1}, 
G. Bonnoli\inst{1}, 
A. Doi\inst{20}, 
L. Maraschi\inst{1}, 
G. Tagliaferri\inst{1}, 
F. Tavecchio\inst{1}, 
}

\institute{
INAF -- Osservatorio Astronomico di Brera, via E. Bianchi 46, 23807, Merate (LC), Italy\\
\email{luigi.foschini@brera.inaf.it}
\and
Max-Planck-Institut f\"ur Radioastronomie, Auf dem H\"ugel 69, 53121 Bonn, Germany
\and
Cahill Center for Astronomy \& Astrophysics, Caltech, 1200 E. California Blvd, Pasadena, CA 91125, USA
\and
Aalto University Mets\"ahovi Radio Observatory, Mets\"ahovintie 114, FIN-02540 Kylm\"al\"a, Finland 
\and
Department of Physics, Purdue University, West Lafayette, IN 47907, USA
\and
Department of Astronomy and Physics, Saint Mary's University, 923 Robie Street, Halifax, NS B3H 3C3, Canada
\and
Department of Natural Sciences, Shawnee State University, 940 2nd Street, Portsmouth, OH 45662, USA
\and
National Astronomical Observatory of Japan, 2-21-1 Osawa, Mitaka, Tokyo, 181-8588, Japan
\and
Crimean Astrophysical Observatory, 98409 Nauchny, Crimea, Ukraine
\and
Pulkovo Observatory, 196140 St. Petersburg, Russia
\and
NASA Goddard Space Flight Center, Greenbelt, MD 20771, USA
\and
ITPA, Universit\"at W\"urzburg, Campus Hubland Nord, Emil-Fischer-Str. 31 D-97074 W\"urzburg, Germany.
\and
ISDC Data Centre for Astrophysics, Chemin d'Ecogia 16, CH-1290, Versoix, Switzerland
\and
Instituto Nacional de Astrof\'isica, \'Optica y Electr\'onica, Tonantzintla, Puebla 72840, Mexico
\and
Department of Astronomy \& Astrophysics, Pennsylvania State University, University Park, PA 16802, USA
\and
INAF -- Istituto di Radioastronomia, via Gobetti 101, 40129, Bologna, Italy
\and
Astro Space Center of the Lebedev Physical Institute, Profsoyuznaya 84/32, 117997 Moscow, Russia
\and
Instituto de Radioastronom\'ia Milim\'etrica (IRAM), Avenida Divina Pastora 7, Local 20, 18012 Granada, Spain
\and
SISSA, via Bonomea 265, 34136 Trieste, Italy
\and
Institute of Space and Astronautical Science, JAXA, 3-1-1 Yoshinodai, Sagamihara, Kanagawa 229-8510, Japan
}

   \date{Received --; accepted --}
 
\abstract{We present more than three years of observations at different frequencies, from radio to high-energy $\gamma$-rays, of the Narrow-Line Seyfert 1 (NLS1) Galaxy PMN~J0948$+$0022 ($z=0.585$). This source is the first NLS1 detected at energies above 100 MeV and therefore can be considered the prototype of this emerging new class of $\gamma$-ray emitting active galactic nuclei (AGN). The observations performed from 2008 August 1 to 2011 December 31 confirmed that PMN~J0948$+$0022 generates a powerful relativistic jet, able to develop an isotropic luminosity at $\gamma$-rays of the order of $10^{48}$~erg~s$^{-1}$, at the level of powerful quasars. The evolution of the radiation emission of this source in 2009 and 2010 followed the canonical expectations of relativistic jets, with correlated multiwavelength variability ($\gamma$-rays followed by radio emission after a few months), but it was difficult to retrieve a similar pattern in the light curves of 2011. The comparison of $\gamma$-ray spectra before and including 2011 data suggested that there was a softening of the high-energy spectral slope. We selected five specific epochs to be studied by modelling the broad-band spectrum, characterised by an outburst at $\gamma$-rays or very low/high flux at other wavelengths. The observed variability can largely be explained either by changes in the injected power, the bulk Lorentz factor of the jet or the electron spectrum. The characteristic time scale of doubling/halving flux ranges from a few days to a few months, depending on the frequency and the sampling rate. The shortest doubling time scale at $\gamma$-rays is $2.3\pm 0.5$ days. These small values underline the need of highly-sampled multiwavelength campaigns to better understand the physics of these sources.
}


\authorrunning{L. Foschini et al.}
\titlerunning{Monitoring of the NLS1 PMN~J0948$+$0022}

\maketitle

\section{Introduction}
PMN~J0948$+$0022 (also known as SDSS~J094857.31$+$002225.4, $z=0.585\pm 0.001$\footnote{Sloan Digital Sky Survey Team, SDSS Data Release 2, Nov. 18, 2004 from: http://www.sdss.org/dr2/products/spectra/getspectra.html}) is a very peculiar AGN, with an optical spectrum of a Narrow-Line Seyfert 1 (NLS1) galaxy (Williams et al. 2002). Zhou et al. (2003) first reported the unexpected connection of its NLS1 character with strong and flat-spectrum radio emission. The optical spectrum of PMN~J0948$+$0022 exhibits the usual characteristics of NLS1s (cf. Osterbrock \& Pogge 1985, Goodrich 1989): FWHM(H$\beta$) $=1500\pm 55$~km/s, [OIII]/H$\beta <3$, strong optical FeII emission (Zhou et al. 2003). Such types of AGN are generally radio-quiet (Ulvestadt et al. 1995, Komossa et al. 2006), so it was a surprise to detect strong radio emission. The radio power at 5~GHz was measured to be $\sim 10^{43}$~erg~s$^{-1}$, a relatively small value for a blazar, but very high for a NLS1, resulting in a radio loudness $RL=f_{\rm 5~GHz}/f_{\rm 440~nm}\gtrsim 10^3$ (Zhou et al. 2003). This value is among the highest for a NLS1 (although not the highest, cf. Fig.~3 in Foschini 2011a). The inverted spectrum ($\alpha_{\rm 5-10~GHz}\sim -0.24$, with $f_{\nu}\propto \nu^{-\alpha}$) and high brightness temperature ($T_{\rm b}\gtrsim 10^{13}$~K) suggest the presence of a relativistic jet viewed at small angles (Zhou et al. 2003, Doi et al. 2006), similar to the conditions found in blazars (Blandford \& Rees 1978).

Confirmation of a powerful relativistic jet viewed at small angle arrived with the launch of the {\it Fermi Gamma-ray Space Telescope} (hereafter {\it Fermi}) carrying onboard the Large Area Telescope (LAT, LAT Coll. 2009a), which operates in the energy band from 100~MeV to more than $300$~GeV and performs an all-sky survey every two orbits ($\sim 3$~hours). During its early months of operation, LAT detected variable high-energy $\gamma$-ray radiation from PMN~J0948$+$0022 (LAT Coll. 2009b,c, 2010), and after one year the number of $\gamma$-NLS1s increased to four (LAT Coll. 2009d). To date there are five high-confidence (Test Statistic $TS>25$, see Mattox et al. 1996 for a definition of $TS$) and three low-confidence ($TS=9-25$) $\gamma$-ray detections of NLS1 plus a similar number of candidates in the possible parent population\footnote{The current status of the search for high-energy $\gamma$-ray emission from NLS1s and the related suggested parent population is available at the web page {\tt http://tinyurl.com/gnls1s} prepared by LF.} (see Foschini 2011a for a review). 

The importance of the research resides in understanding a few fundamental points. On one side, there are blazars (beamed population) plus radio galaxies (unbeamed parent population): they have the mass of the compact object in the range $\sim 10^{8}-10^{10}M_{\odot}$, accretion discs spanning a wide range of luminosities, from $\sim 10^{-6}$ to a significant fraction of the Eddington limit (e.g. Ghisellini et al. 2010, Foschini 2011b), and the host galaxy is elliptical (e.g. Kirhakos et al. 1999). At the other side, there are the $\gamma$-NLS1s (beamed population) plus an unbeamed parent population yet to be exploited, which have the central supermassive black hole with masses between $\sim 10^{6}-10^{8}M_{\odot}$, accretion disc luminosities close to the Eddington limit (e.g. Foschini 2011a,b), and a few of them are hosted in spirals possibly having undergone a recent merger (see Yuan et al. 2008, Foschini 2011a,b, 2012, Hamilton \& Foschini 2012).  Therefore, it is now possible to study AGN with powerful relativistic jets in objects that exhibit an unexplored range of central black hole masses and rates (cf Fig. 8 in Foschini 2011a).

PMN~J0948$+$0022 has been highly active during the past three and a half years. Radio data from Owens Valley Radio Observatory (OVRO) extending back to one year before the launch of {\it Fermi} (see LAT Coll. 2009b, Fig.~2, Panel C) displayed a radio flux density at 15~GHz that reached up to $\sim 1$~Jy on 2008 January, a factor $\sim 2-3$ above the value at the time of the discovery (2008 August-December). The first multiwavelength (MW) campaign organised in 2009 March-July revealed continuing activity (LAT Coll. 2009c, see also Giroletti et al. 2011 for e-VLBI observations) and in 2010 July the source erupted into an exceptional outburst with a peak $\gamma$-ray isotropic luminosity of $\sim 10^{48}$~erg~s$^{-1}$ (LAT Coll. 2010b, Foschini 2010, Foschini et al. 2011a,b). The new data presented here show that the activity continued into 2011 as well. 

PMN~J0948$+$0022 has been the target of several observations at almost every waveband during the past three and a half years (2008 August 1 - 2011 December 31), resulting in an impressive wealth of data. We present here a comprehensive study of all these data in order to set up some firm pillars in our understanding of the prototype of this new class of $\gamma$-ray AGN. Part of the data presented here has been already published by the LAT Coll. (2009b,c,d), Foschini et al. (2011a,b), Angelakis et al. (2012), but they have been reanalysed by using the most recent software version and calibration database. This is particularly important in the case of LAT data, where the change from the P6 to P7 instrument response function determined important changes in the analyses (cf. LAT Coll. 2010c with LAT Coll. 2012). The effects of these changes on the 2010 July outburst have been presented in Foschini et al. (2011b), and the present work reports about the reanalysis of the whole period 2008-2011. In addition, the 2011 data are presented here for the first time.

The work is structured as follows: after the introduction, Sect.~2 deals with the descriptions of the individual instruments and facilities used for the observations, and the techniques for data analysis. A general overview of the data (avoiding interpretations) is given in Sect.~3. Five specific periods have been selected and the corresponding broad-band spectra have been studied with the one-zone synchrotron and inverse-Compton model by Ghisellini \& Tavecchio (2009). This part is described in Sect.~4. Some final remarks (Sect.~5) conclude the work. 

Throughout this work, the luminosities have been calculated by assuming a $\Lambda$CDM cosmology with a Hubble-Lema\^itre constant $H_{0}=70$~km~s$^{-1}$~Mpc$^{-1}$ and $\Omega_{\Lambda}=0.73$ (Komatsu et al. 2011). The luminosity distance of PMN~J0948$+$0022 is then 3462~Mpc and 1 arcsec corresponds to 6.68~kpc. In the following, the dates will be indicated either with calendar scheme, or Modified Julian Date (MJD), or days since 2008 August 1 (MJD$-$54679).

\section{Data Analysis}

\subsection{Fermi Large Area Telescope}
LAT photons of class 2 (``source'' type) with energy between 100 MeV and 100 GeV recorded between 2008 August 4 15:43 UTC and 2011 December 31 24:00 UTC in a circular region around the position of PMN~J0948$+$0022 with $10^{\circ}$ radius were retrieved from the {\it Fermi} Science Support Center\footnote{{\tt http://fermi.gsfc.nasa.gov/ssc/data/}}. The data were analysed by using {\tt LAT Science Tools v. 9.27.1}, together with the Instrument Response Function (IRF) Pass 7 and the corresponding isotropic and Galactic diffuse background models\footnote{See the documentation available online at the {\it Fermi} SSC.}. The adopted procedures are standard and described in detail in the analysis threads available online at the {\it Fermi} SSC. 

The data were fitted with a single power-law model. The analysis of all the data integrated into one single block resulted in 4951 photons predicted by the model, which in turn corresponds to a flux $F_{0.1-100~\rm GeV}=(1.36\pm 0.03)\times 10^{-7}$~ph~cm$^{-2}$~s$^{-1}$ and a photon index $\Gamma = 2.67\pm 0.03$. The likelihood of detection is expressed by means of the Test Statistic ($TS$, Mattox et al. 1996), which, as a rule of thumb, is linked to the significance $n\sigma$ as $\sqrt{TS}\sigma$. For the {\it Fermi} observations of PMN~J0948$+$0022, $TS=2015$. 

In the 2FGL catalogue (LAT Coll. 2012), the energy distribution $N(E)$ of PMN~J0948$+$0022 is fitted with a log-parabola model of the form:

\begin{equation}
N(E) = N_0 \left (\frac{E}{E_{\rm p}} \right)^{\alpha + \beta \ln (\frac{E}{E_{\rm p}})}
\end{equation}

\noindent where $N_{0}$ is the normalisation [ph~cm$^{-2}$~s$^{-1}$~MeV$^{-1}$], $\alpha$ is the photon index at the pivot energy $E_{\rm p}$ [MeV], and $\beta$ is the curvature index. Specifically for PMN~J0948$+$0022, $N_{0}=(1.39\pm 0.07)\times 10^{-10}$~ph~cm$^{-2}$~s$^{-1}$~MeV$^{-1}$, $\alpha=2.26\pm 0.08$, $\beta=0.26\pm 0.06$, and $E_{\rm p}\sim 272$~MeV and refer to the period 2008 August 4 -- 2010 July 31 (LAT Coll. 2012).

For comparison, using the same model for the integrated data considered here (2008 August 4 -- 2011 December 31) we obtain values of $N_{0}=(1.23\pm 0.03)\times 10^{-10}$~ph~cm$^{-2}$~s$^{-1}$~MeV$^{-1}$, $\alpha=2.45\pm 0.03$, $\beta=0.23\pm 0.02$, and $E_{\rm p}=(313\pm 3)$~MeV. In total, 4802 photons are predicted by the model, corresponding to an integrated flux of $F_{0.1-100~\rm GeV}=(1.21\pm 0.03)\times 10^{-7}$~ph~cm$^{-2}$~s$^{-1}$, and $TS=2070$. The changes are basically due to the addition of the 2011 data, when the source was strongly active for almost the entire year (see Section 3, Fig.~\ref{Fig:curve}, {\it top panel}), and slight softening of the spectrum occurred ($\alpha$ increased by $\sim 0.2$, while the curvature $\beta$ remained constant). 

\begin{figure}[!t]
\centering
\includegraphics[angle=270,scale=0.35]{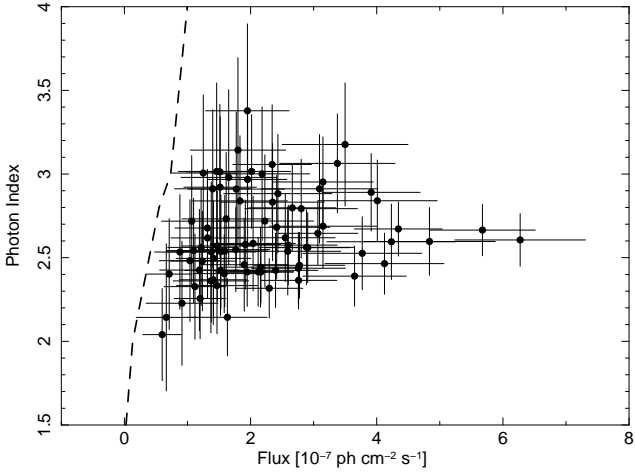}
\caption{$\gamma$-ray power-law model photon index as a function of the $0.1-100$~GeV flux. The data points represent 7-day bins. The dashed line marks the {\it Fermi}/LAT 7-day detection threshold ($TS=9$) as extrapolated from LAT Coll. (2010c).}
\label{Fig:GAMMA}
\end{figure}

Fig.~\ref{Fig:GAMMA} displays the behaviour of the $\gamma$-ray spectral slope as a function of the 7-day integrated flux in the $0.1-100$~GeV energy band. The measured photon index is that of the power-law model and no strong spectral changes can be observed on short time scales, while there could be on longer periods, as shown from the above comparison of the 2FGL data with the present work. 

The $\gamma$-ray light curve in Fig.~\ref{Fig:curve} ({\it top panel}) and \ref{Fig:CURVE1} ({\it top left panel}) was built with one-day time bins, assuming a power-law model with $\Gamma=2.67$ (i.e. the value from the integration of the whole data set), and with a detection threshold of $TS=9$. Since the source is a confirmed $\gamma$-ray emitter, we relaxed the statistical requirements for valid detections that has been adopted in recent works (e.g. Foschini et al. 2011a,b). Due to its scanning mode, {\it Fermi} monitors any source in the sky almost continuously. Sometimes, PMN~J0948$+$0022 was not detected on daily time-scale, but for clarity we do not plot upper limits. The $TS=4$ average upper limit for one-day exposure ($\sim 2.3\times 10^{-7}$~ph~cm$^{-2}$~s$^{-1}$, by fixing $\Gamma=2.67$) is indicated in Fig.~\ref{Fig:curve} ({\it top panel}) and \ref{Fig:CURVE1} ({\it top left panel}) with an horizontal dotted line.

\subsection{Swift}
The data of PMN~J0948$+$0022 from the three {\it Swift} instruments, the Burst Alert Telescope (BAT, Barthelmy et al. 2005), the X-Ray Telescope (XRT, Burrows et al. 2005), and the UltraViolet Optical Telescope (UVOT, Roming et al. 2005) were downloaded from the HEASARC archive and analysed with the {\it Swift} tools included in the {\tt HEASoft v. 6.12} package\footnote{Including the XRT Data Analysis Software (XRTDAS) developed under the responsibility of the ASI Science Data Center (ASDC), Italy.}, together with the {\tt CALDB} updated on 2012 March 22. 

{\it Swift} observed the source for the first time on 2008 December 5 soon after the discovery of the $\gamma$-ray emission (LAT Coll. 2009b). During the 2009 MW campaign, there were 11 snapshots (LAT Coll. 2009c) and one a few days before the outburst of 2010 July (Foschini et al. 2011a). Ten more exposures were performed in 2011, as part of a monitoring program linked to the Effelsberg radio observations. In total, from the period 2008-2011, there are 23 observations with XRT exposures from 1 to 5~ks each.

All the available pointed BAT observations were combined to search for a possible detection. In total, $\sim 86$~ks of data were accumulated, but no detection was made. The upper limits at 99.97\% confidence level were $1.7\times 10^{-10}$~erg~cm$^{-2}$~s$^{-1}$ in the $20-40$~keV energy band and $2.3\times 10^{-10}$~erg~cm$^{-2}$~s$^{-1}$ in the $40-100$~keV band. Cusumano et al. (2010) reported a detection by integrating 54 months of BAT data (thus including also the {\it Swift} observations when PMN~J0948$+$0022 was not on axis, but is still within the BAT field of view), with a flux of $(1.2\pm 0.8)\times 10^{-11}$~erg~cm$^{-2}$~s$^{-1}$ in the $10-150$~keV energy band\footnote{It is worth noting that the 58-month survey by Baumgartner et al. does not report this detection. See: {\tt http://swift.gsfc.nasa.gov/docs/swift/results/bs58mon/}}. Despite the large error, the flux seems to be consistent with what is expected from the modelling of the spectral energy distribution (see Section 4, Fig.~\ref{Fig:SED}).  

\begin{figure}[!t]
\centering
\includegraphics[angle=270,scale=0.35]{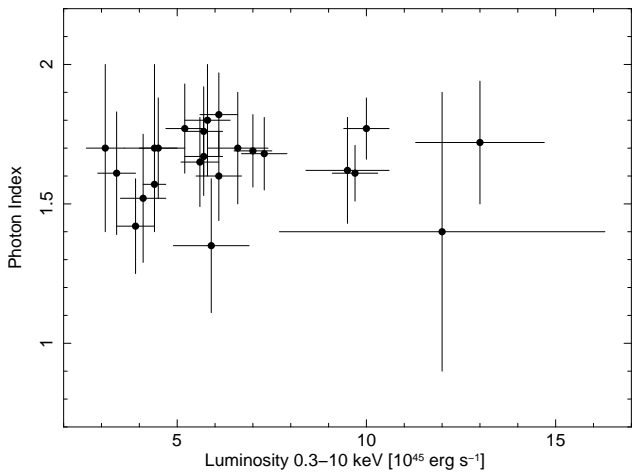}
\caption{X-ray power-law model photon index plotted as a function of the 0.3-10~keV luminosity. No obvious luminosity dependent spectral changes are seen.}
\label{Fig:XRAY}
\end{figure}

XRT was set to work in photon counting mode (Hill et al. 2004) and we analysed all the single-to-quadruple events (grades 0-12). The extracted spectra were rebinned to have at least 20 counts per bin to apply the $\chi^2$ statistics. When this was not possible, we evaluated the likelihood using Cash statistics (Cash 1979). The data were fitted with a redshifted power-law model with Galactic absorption ($N_{\rm H}=5.23\times 10^{20}$~cm$^{-2}$, Kalberla et al. 2005). The individual snapshot does not show significant spectral changes as a function of the emitted power, but the statistics are not sufficient to test if more complex models can be employed (Fig.~\ref{Fig:XRAY}). When all the X-ray data are integrated (the resulting global exposure is $81.6$~ks), the spectrum is best fitted ($\chi^2=241.66$, $dof=200$) with $\Gamma=1.67\pm 0.03$ and normalisation at 1 keV equal to $(1.50\pm 0.05)\times 10^{-3}$~ph~cm$^{-2}$~s$^{-1}$~keV$^{-1}$. The integrated observed flux in the $0.3-10$~keV band is $(4.5\pm 0.2)\times 10^{-12}$~erg~cm$^{-2}$~s$^{-1}$. The modelling of the integrated spectrum can be improved ($\chi^2=210.54$, $dof=198$, F-test $>99.99$\%) by adopting a broken power-law model with the following parameters: $\Gamma_1=1.91\pm0.10$, $\Gamma_2=1.56_{-0.07}^{+0.05}$, $E_{\rm break}=1.22_{-0.19}^{+0.51}$~keV, and normalisation $(6.7\pm 0.3)\times 10^{-4}$~ph~cm$^{-2}$~s$^{-1}$~keV$^{-1}$ at 1 keV.

It is possible to model the spectrum also by using log-parabolic models ({\tt logpar} in the {\tt xspec} syntax), which have been applied to blazars, specifically to High-Frequency-Peaked BL Lac Objects (e.g. Massaro et al. 2004, Foschini et al. 2007). The best-fit parameters are $\alpha=1.79\pm 0.04$ and $\beta=-0.27\pm 0.04$, with the pivot energy fixed at 1~keV. The $\chi^2$ test resulted in a value of 210.34 for 199 degrees of freedom, which in turn translates into a {\it F-test} better than $99.99$\% with respect to the single power-law model. A variant of the above model, {\tt eplogpar} in {\tt xspec} (see Tramacere et al. 2007), resulted in similar values for the $\chi^2$ and {\it F-} tests ($\chi^2=210.46$, $dof=199$, $F-test>99.99$\%) and the following model parameters: the peak energy in $\nu F_{\nu}$ is $2.4\pm 0.2$ keV and the curvature $\beta=-0.28\pm 0.09$. In both cases, the improvement with respect to the single power-law model is significant, but there is no preference with respect to the broken power-law model. Basically, the spectrum indicates a change in the slope around $1-2$ keV, with the high-energy photon index harder than that at lower energies. 

UVOT data were taken with all six filters ($V$, $B$, $U$, $UVW1$, $UVM2$, $UVW2$) for most the observations. They were analysed by extracting the source counts from a $5''$-sized region and the background from a source-free annulus centred on the source, and with internal and external dimensions of $7''$ and $40''$, respectively. The observed magnitudes were dereddened by using $A_{\rm V}=0.28$ and the extinction laws by Cardelli et al. (1989). The intrinsic magnitudes were then converted in physical units by using the zero-points and conversion factors of the {\it Swift} {\tt CALDB} (Poole et al. 2008, Breeveld et al. 2010).

\subsection{Instituto Nacional de Astrof\'isica, \'Optica y Electr\'onica (INAOE)}
PMN J0948$+$0022 was observed at near-infrared (NIR) wavelengths at the $2.1$~m telescope ``Guillermo Haro'' (Cananea, Sonora, Mexico). The NIR camera ``CANICA'' has a Rockwell $1024 \times 1024$ pixel Hawaii infrared array ($0.32$ arcsec/pix), operating at $75.4$~K, with standard  J (1.164 - 1.328 $\mu$m), H (1.485 - 1.781 $\mu$m) and K$_{\rm s}$ (1.944 - 2.294 $\mu$m) filters in place. Observations were carried out in series of $10$ dithered frames in each filter, with a proper number of additional observations for the K$_s$ filter. Data sets were corrected for bias and flat-fielding, the latter obtained from sky frames derived from the dithered ones, and coadded.

\subsection{Effelsberg and IRAM}
The cm/mm radio light curves of PMN J0948+0022 have been obtained within the framework of a {\it Fermi} related monitoring program of $\gamma$-ray blazars (F-GAMMA program, Fuhrmann et al. 2007, Angelakis et al. 2008) as well as a dedicated NLS1 monitoring program. The millimetre observations are closely coordinated with the more general flux density monitoring conducted by IRAM (Instituto de Radioastronom\'ia Milim\'etrica), and data from both programs are included in this paper. The overall frequency range spans from 2.64\,GHz to 142\,GHz using the Effelsberg 100-m and IRAM 30-m (at Pico Veleta) telescopes.

The Effelsberg measurements were conducted with the secondary focus heterodyne receivers at 2.64, 4.85, 8.35, 10.45, 14.60, 23.05, 32.00, and 43.00\,GHz.  The observations were performed quasi-simultaneously with cross-scans, that is slewing over the source position, in azimuth and elevation direction with adaptive number of sub-scans for reaching the desired sensitivity (for details, see Fuhrmann et al. 2008; Angelakis et al.  2008). Consequently, pointing off-set correction, gain correction, atmospheric opacity correction and sensitivity correction have been applied to the data.

The IRAM 30-m observations were carried out with calibrated cross-scans using the `ABCD' SIS (until March 2009) and new EMIR horizontal and vertical polarisation receivers operating at 86.2 and 142.3\,GHz. The opacity corrected intensities were converted into the standard temperature scale and finally corrected for small remaining pointing offsets and systematic gain-elevation effects. The conversion to the standard flux density scale was done using the instantaneous conversion factors derived from frequently observed primary (Mars, Uranus) and secondary (W3(OH), K3-50A, NGC\,7027) calibrators.

The complete data set has been published in Angelakis et al. (2012), together with data of other NLS1s. The brightness temperature of PMN J0948$+$0022 was measured as $8\times 10^{12}$~K at 4.85~GHz, $2\times 10^{12}$~K at 14.6~GHz, and $1.5\times 10^{11}$~K at 32 GHz, which can be translated into requirement of Doppler factors greater than 6, 4, and 2, respectively (Angelakis et al. 2012).

\begin{figure*}[!t]
\centering
\includegraphics[scale=0.9]{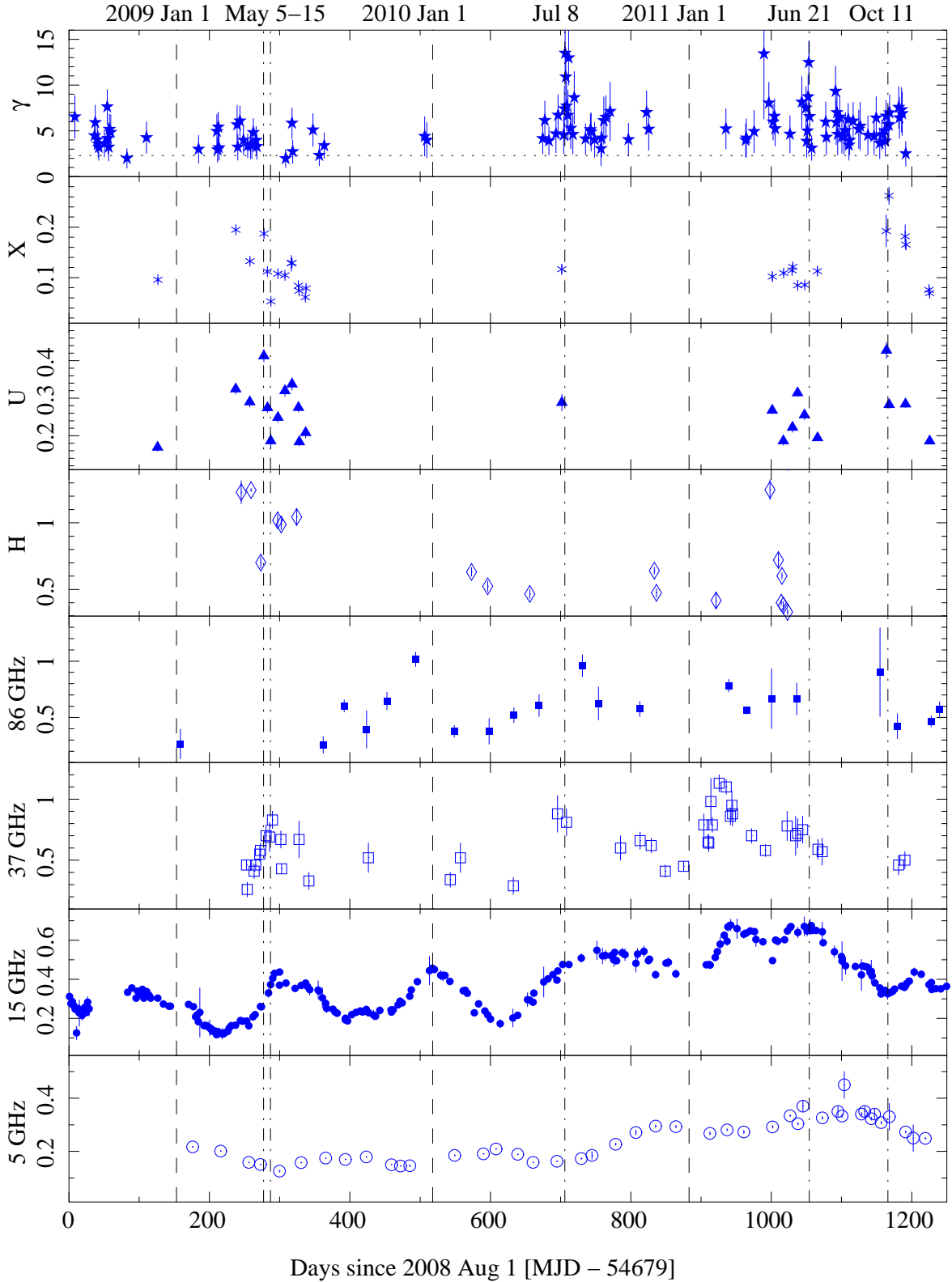}
\caption{Light curves at various frequencies. From top to bottom: $\gamma$-rays $0.1-100$~GeV from {\it Fermi}/LAT with 1 day time bin [$10^{-7}$~ph~cm$^{-2}$~s$^{-1}$]; X-ray $0.3-10$~keV from {\it Swift}/XRT [c~s$^{-1}$]; U filter (350~nm) from {\it Swift}/UVOT [mJy]; H filter ($1.65$~$\mu$m) from INAOE [mJy]; 86~GHz from IRAM [Jy]; 37~GHz from Mets\"ahovi [Jy]; 15~GHz from OVRO, Effelsberg, and MOJAVE [Jy]; 5~GHz from Effelsberg and Medicina. Time starts on 2008 August 1 00:00 UTC (MJD 54679). {\it Fermi} performs an all-sky survey every two orbits (3 hours) and therefore it observed PMN~J0948$+$0022 almost continuously. However, for clarity, we do not plot the upper limits. The $TS=4$ average sensitivity of LAT for one day exposure is indicated with the horizontal dotted line in the top panel, which corresponds to $\sim 2.3\times 10^{-7}$~ph~cm$^{-2}$~s$^{-1}$ (assuming $\Gamma = 2.67$).
Vertical dashed lines are shown at the beginning of each year (Jan 1). Vertical dot-dashed lines indicated the days referring to the spectral energy distributions selected as example (see Sect.~4).}
\label{Fig:curve}
\end{figure*}

\subsection{Mets\"ahovi}
The Mets\"ahovi radio telescope is a radome enclosed paraboloid antenna ($13.7$~m) situated in Finland and operating at 37~GHz. The detection limit (signal-to-noise ratio $>4$) of the telescope at 37~GHz is of the order of 0.2~Jy under optimal conditions and with a typical integration time of $1200-1400$~s. The flux density scale was set by observations of DR 21, while 3C 84, 3C 274, and NGC 7027 were used as secondary calibrators. More details about data reduction and analysis can be found in Ter\"asranta et al. (1998). The error in the flux density takes into account the background and the absolute calibration.

\subsection{Owens Valley Radio Observatory}
Regular 15 GHz observations of PMN~J0948$+$0022 were carried out as part of a high-cadence $\gamma$-ray blazar monitoring program using the OVRO 40~m telescope (Richards et al. 2011). This program, which commenced in late 2007, now includes about 1600 sources, each observed with a nominal twice per week cadence.

The OVRO 40~m uses off-axis dual-beam optics and a cryogenic high electron mobility transistor (HEMT) low-noise amplifier with a 15.0~GHz center frequency and 3~GHz bandwidth. The total system noise temperature is about 52~K, including receiver, atmosphere, ground, and CMB contributions. The two sky beams are Dicke switched using the off-source beam as a reference, and the source is alternated between the two beams in an ON-ON fashion to remove atmospheric and ground contamination. A noise level of approximately 3--4~mJy in quadrature with about 2\% additional uncertainty, mostly due to pointing errors, is achieved in a 70~s integration period. Calibration is achieved using a temperature-stable diode noise source to remove receiver gain drifts and the flux density scale is derived from observations of 3C~286 assuming the Baars et al. (1977) value of 3.44~Jy at 15.0~GHz. The systematic uncertainty of about 5\% in the flux density scale is not included in the error bars.  Complete details of the reduction and calibration procedure are found in Richards et al. (2011).

\subsection{VLBA 2~cm (MOJAVE Program)}
The MOJAVE Program (Monitoring Of Jets in Active galactic nuclei with VLBA Experiments, Lister et al. 2009) is a high-resolution high-dynamic range (about 8000:1) survey at 15.4~GHz performed with the Very Large Baseline Array (VLBA). It counts about 300 active galactic nuclei in the Northern hemisphere, which are observed on a roughly regular basis. Presently, the program includes four radio-loud NLS1s detected at $\gamma$-rays (J0324$+$3410, J0849$+$5108, J0948$+$0022, J1505$+$0326). Specifically, PMN J0948$+$0022 has been observed 11 times between 2009 May 28, and 2011 December 12 (Table~\ref{table:mojave}). Details of data processing are available in Lister et al. (2009) and Lister \& Homan (2005). The analysis of the 2009 and part of 2010 data resulted in a brightness temperature of the core in excess of $6\times 10^{12}$~K, a jet apparent opening angle of $21^{\circ}$, a jet position angle of $24^{\circ}$, an average fractional polarisation of $0.8$\%, and an average Electric Vector Position Angle (EVPA) of $142^{\circ}$ (Lister et al. 2011). There is negligible difference between the single dish (OVRO) and MOJAVE (VLBA) 2 cm flux densities at all epochs, indicating no bright arcsec scale radio structure (Lister et al. 2011).

The three initial epochs\footnote{Images with animations can be found at:\\
{\tt http://tinyurl.com/mojave0948p0022}} showed two moving features, but these faded rapidly and thus their measured speeds cannot be considered reliable.

\begin{table}[!h]
\caption{VLBA observations performed within the MOJAVE project.}
\begin{center}
\begin{tabular}{lccc}
\hline
\hline
Date       & Intensity [mJy] & Polarisation [\%] & EVPA [deg] \\
\hline
2009 May 28 & 384 & 0.7 & 131\\
2009 Jul 23 & 462 & 1.2 & 146\\
2009 Dec 10 & 675 & 0.5 & -\\
2010 Sep 17 & 669 & 3.7 & 51\\
2010 Nov 04 & 627 & 1.8 & 49\\
2010 Nov 29 & 486 & 1.6 & 49\\
2011 Feb 20 & 498 & 1.0 & 19\\
2011 May 26 & 540 & 1.8 & 49\\
2011 Jun 24 & 389 & 2.1 & 51\\
2011 Sep 12 & 343 & 1.3 & 40\\
2011 Dec 12 & 439 & 0.8 & 77\\
\hline
\hline
\end{tabular}
\end{center}
\label{table:mojave}
\end{table}

\subsection{Medicina}
We observed PMN\,0948+0022 with the 32m Medicina radio telescope eight times between 2011 June and November, at a frequency of 5 and/or 8.4 GHz. At each epoch and frequency, we performed $\sim 10$ cross scans on the source using the new Enhanced Single-dish Control System (ESCS) acquisition system. The typical on source integrated time is $\sim 1$ minute per band. Since the signal-to-noise ratio in each scan across the source was low (typically $\sim 3$), we performed a stacking analysis of the scans, which allowed us to significantly improve the signal-to-noise ratio and the accuracy of the measurement. Finally, we calibrated the flux density with respect to simultaneous observations of 3C\,286.

\subsection{Sloan Digital Sky Survey (SDSS)}
We also reanalysed the non-contemporaneous data of the SDSS. The images were taken in 1999 March, and the spectrum (covering $3800-9200$ \AA, or rest-frame $2400-5600$ \AA) in 2000 March.  The host galaxy is unresolved in the SDSS images, which is not surprising, given its redshift ($z=0.5846$).  The spectrum shows evidence for a young stellar population, in addition to the nuclear source (Hamilton \& Foschini 2012). High-resolution observations (e.g. with {\it Hubble Space Telescope}) are required to study the morphology of the host galaxy.

\section{Overview of the data}
All the data available for the present study are displayed in the figures of Appendix A, while a sub-sample with the most representative light curves is shown in Fig.~\ref{Fig:curve}. The period covered is from 2008 August 1 to 2011 December 31. Both samples clearly show a high degree of activity in PMN~J0948$+$0022 at all wavelengths. The most complete coverage is at $\gamma$-rays with {\it Fermi}/LAT (daily), and at 15~GHz with OVRO (twice per week), plus contributions from Effelsberg (once per month) and the MOJAVE Program (Table~\ref{table:mojave}). At the other wavelengths the data are basically clustered around two main MW campaigns organised in 2009 (LAT Coll. 2009c) and 2011 (new data).

\subsection{2008}
As already published in LAT Coll. (2009b), radio observations at OVRO (15 GHz) before the launch of {\it Fermi} indicated a strong radio emission with a peak of $0.81$~Jy on 2008 Jan 9 (MJD $54474$). At the start of the scientific operations of the LAT onboard {\it Fermi}, PMN~J0948$+$0022 was detected at a level of a few $\times 10^{-7}$ ph~cm$^{-2}$~s$^{-1}$ making it the first NLS1 detected at high-energy $\gamma$-rays (LAT Coll. 2009b, 2010a).  

\subsection{2009}
We confirm the findings of the 2009 MW Campaign, performed between 2009 Mar 20 and Jul 5 (LAT Coll. 2009c). The source displayed some activity at $\gamma$-rays in April, with the most significant (i.e. with greatest $TS$) peak detected on April 1 (MJD 54922; day 243 in Fig.~\ref{Fig:curve}) with a flux of $(4.2\pm 1.5)\times 10^{-7}$ ph~cm$^{-2}$~s$^{-1}$ in the $0.1-100$ GeV energy band with a photon index $\Gamma = 2.1\pm 0.2$ ($TS=43$). Other detections on daily basis were recorded by {\it Fermi}/LAT during almost the whole month, with fluxes of a few times $10^{-7}$ ph~cm$^{-2}$~s$^{-1}$ ($0.1-100$ GeV). Near-Infrared observations at INAOE show a relatively bright state, with observed magnitudes $J=15.87\pm 0.05$, $H=14.90\pm 0.07$, $K_{\rm s}=13.99\pm 0.09$, while optical-polarimetric observations (V filter) performed in the period 2009 Mar 30 - Apr 10 at the KANATA telescope resulted in a $19$\% polarisation fraction (Ikejiri et al. 2011).

Radio observations in 2009 April reported low flux densities of the order of $\sim 0.2$~Jy at 15~GHz (see Fig.~\ref{Fig:curve}), but increasing and peaking in mid-end May. At 15 GHz, with the densest sampling, the peak of $0.43$~Jy was measured on 2009 May 21 (MJD 54972, day 293 in figures), that is 50 days after the $\gamma$-ray peak. Mets\"ahovi (37~GHz) reported $0.83$~Jy on the evening of 2009 May 17 (MJD 54968.72, day $\sim 290$). Instead, on 2009 May 27 (MJD 54978.77 or day $\sim 300$), Effelsberg measured flux densities reach $0.269$~Jy at $10.45$~GHz), and $0.189$~Jy at $8.35$~GHz (Fig.~\ref{Fig:CURVE3}). 

The {\it Swift} optical-to-X-ray monitoring was much coarsely sampled. However, two observations performed on 2009 May 5 and 15 show a drop in the $0.3-10$~keV flux from $(7.5\pm 0.3)\times 10^{-12}$~erg~cm$^{-2}$~s$^{-1}$ to $(2.1\pm 0.4)\times 10^{-12}$~erg~cm$^{-2}$~s$^{-1}$, corresponding to an intrinsic (i.e. redshift corrected) halving time scale of $3.4\pm 0.7$~day. At the same time, the UVOT spectrum exhibited a slope change. Over a 5-day period centred on 2009 May 5, when the $\gamma$-ray flux was at level of $(2.2\pm0.4)\times 10^{-7}$ ph~cm$^{-2}$~s$^{-1}$ ($TS=10$), the optical/UV spectrum was rather flat. Ten days later the $\gamma$-ray flux fell below the LAT detectability ($<8\times 10^{-8}$ ph~cm$^{-2}$~s$^{-1}$, upper limit at $TS=4$ with 5 days integration). At the same time, a flux density drop of $\sim 2.8$ was seen in optical filters, whereas the flux density decreased by only $\sim 1.8$ in the UV.  Interestingly, the $H$ filter detected a flux density drop on 2009 May 1 (MJD 54952, day 273), prior to the optical-to-X-ray fluxes. The NIR flux density had decreased by roughly a factor 2 since the measurements performed in 2009 April. 

The source ceased detectable $\gamma$-ray activity from 2009 August to about 2010 June, with the exception of a couple of detections in 2009 mid December. There was no information at optical-to-X-rays, but the flux density at radio frequencies was high as measured at 86 GHz (IRAM, more than 1 Jy on 2009 Dec 7) and at 15 GHz (0.46 Jy on 2009 Dec 31). Interestingly, the VLBA 2-cm observation performed on 2009 Dec 10 (MOJAVE), resulted in the lowest polarisation fraction ever measured (0.5\%, Table~\ref{table:mojave}). It was not possible to measure the EVPA in that epoch, but the following observation performed on 2010 Sept 17, revealed that it had changed significantly, rotating by about $90^{\circ}$, from $146^{\circ}$ on 2009 Jul 23, to $51^{\circ}$ (Table~\ref{table:mojave}).

\subsection{2010}
Then, on 2010 July 8 (day 706), PMN J0948$+$0022 exploded in its first known $\gamma$-ray outburst, with isotropic luminosity at $\gamma$-rays in excess of $10^{48}$~erg~s$^{-1}$, comparable with powerful quasars (LAT Coll. 2010b, Foschini 2010, Foschini et al. 2011a,b). At that time, it was not possible to set up an efficient campaign, because of the close apparent position of the Sun, but the data available show a relatively high level of optical-to-X-ray fluxes. Measurements at radio frequencies show an increasing flux density that reached the peak at 15 GHz about 46 days after the first $\gamma$-ray outburst (0.55~Jy on 2010 Aug 22, day 752) and remaining almost constant at this level for about one hundred days. On 2010 Sept 17 (day 777), the VLBA 2-cm observation reported the highest fractional polarisation measured to date at this wavelength (3.7\%, Table~\ref{table:mojave}). The flux density at 37~GHz had a local maximum on June 27 (0.88~Jy, day 695) and July 10 (0.81~Jy, day 708). The next measurement was only on Sept 2 with 0.5~Jy (day 762). The 86 GHz flux density was again at $\sim 1$~Jy level on Aug 2 (day 731).

\subsection{2011}
The year 2011 data are characterised by a prolonged activity both at $\gamma$-rays and radio frequencies. Specifically, Effelsberg and Medicina reported a 5~GHz flux density almost double the values seen in 2008-2009. Starting from 2011 Feb 1 (day 914), Mets\"ahovi (37 GHz) registered a flux density increase, reaching the maximum value of the period, with a value of $1.13$~Jy on 2011 Feb 13 (day 926). During this period there was a further increase in  the 15 GHz flux density at level of $\sim 0.6$~Jy. Instead, the activity at $\gamma$-rays was sparse and of the order of a few times $10^{-7}$ ph~cm$^{-2}$~s$^{-1}$. NIR observations on April 26 (day 998) indicated a high level ($\sim 1.2$~mJy) that dropped by a factor 3 within a couple of weeks. About ten days earlier (day 989) there was some high-level activity at $\gamma$-rays with $F_{0.1-100\rm GeV}=(1.3\pm 0.7)\times 10^{-6}$~ph~cm$^{-2}$~s$^{-1}$.
 
In the optical, the U flux density increased up to $0.3$~mJy (day 1037, Jun 4) and then decreased again. On Jun 13 (day 1046), the 15 GHz flux density reached the local maximum of $\sim 0.7$~Jy and a few days later, on Jun 20 (day 1053), the source erupted to another strong $\gamma$-ray outburst ($L_{\gamma}\sim 10^{48}$~erg~s$^{-1}$). This event was reported by the {\it AGILE} satellite (AGILE Coll. 2011). Another $\gamma$-ray outbursts at a level of $\sim 10^{-6}$~ph~cm$^{-2}$~s$^{-1}$ occurred on Jul 28 (day 1091).

An interesting event occurred on October 9-12 (days 1164-1167). {\it Swift} observed the highest optical/X-ray fluxes ever recorded ($F_{0.3-10\,\rm keV}\sim1.0\times 10^{-11}$~erg~cm$^{-2}$~s$^{-1}$; $F_{\rm U}\sim 0.43$~mJy), but with no corresponding high activity at $\gamma$-rays (``orphan flare''). High flux density was recorded by IRAM at 86 GHz as well, while at lower frequencies (37 and 15 GHz) there was a decreasing trend. VLBA 2-cm observations at the end of 2011 indicated a decrease of the polarisation and some hint for another swing in the EVPA.

\subsection{Time variability}
To obtain an estimate of the shortest observed variability in each band, we follow the method applied in, e.g., Foschini et al. (2011c). We calculated the observed time scale $\tau$ for doubling/halving flux according to the formula:

\begin{equation}
F(t) = F(t_0)\cdot 2^{-(t-t_0)/\tau}
\end{equation}

\noindent where $F(t)$ and $F(t_0)$ are the fluxes at the time $t$ and $t_0$, respectively, of two adjacent points. Although, it is more common to find in the literature the time scale for an exponential flux change (e.g. Scargle 1981, LAT Coll. 2010e), we prefer to adopt the doubling/halving $\tau$ because of the large error bars in the $\gamma$-ray curve. The error in the evaluation of $\tau$ has been calculated according to the bootstrap method, by assuming the worst case ($\tau_{\rm max}$) calculated by using the smallest flux difference and the longest time interval. The results are displayed in Table~\ref{variability} and includes the shortest $\tau$ and also the flux ratio measured between the maximum and minimum of the selected curve. It is worth noting that the displayed $\tau$ is the observed value. The intrinsic characteristic time scale can be obtained by dividing $\tau$ by $(1+z)$.

We note some peculiar variations in our estimates in Table~\ref{variability}, for example the wide variation between neighboring frequencies at 37~GHz and 32~GHz. While such a sensitivity of variability rates with frequency would be quite interesting, it is likely that this is an artifact of the unequal sampling. Effelsberg almost regularly sampled the source with one-month cadence, while other curves, with more irregular sampling (e.g. 37~GHz), made it possible to find also shorter variability ($\sim 2$ days), depending on the clustering of observations around specific events. The most reliable results are those at $\gamma$-rays and at 15~GHz, which are also the curves most densely sampled. In these cases, it is possible to find variability on a few days time scale. The detection of such a day-scale variability at radio frequencies is intriguing, but not extraordinary, since it was already detected in other blazars (cf. Wagner \& Witzel 1995).

The $\gamma$-rays show the most extreme flux variations, with a time scale of a few days, although during outbursts -- when the statistics are sufficient to perform an analysis with shorter time bins -- it is possible to find intraday variability (Foschini et al. 2011a). The relativistic jets of blazars are known to display similar time scales (cf. LAT Coll. 2010e), and during outbursts it is possible to measure variability on time scales of hours (Foschini et al. 2011c). The lowest flux changes are in the UV, which is expected to be dominated by radiation from the accretion disk as in the quasar subclass of blazars (cf Bonning et al. 2012), and at 2.64~GHz, which should refer to the largest spatial scales. Flux changes by factors 5-6 occur at X-rays, infrared H filter, 142 GHz, and 15 GHz. To summarize, the observed variability (amplitude and time scale), by taking into account the bias of the sparse sampling, is consistent with that displayed by flat-spectrum radio quasars.

\begin{table}[!t]
\caption{Variability at different wavelengths. 
Col. [1]: Band/Filter/Frequency of the data; 
Col. [2]: Observed characteristic time-scale for doubling/halving the flux [day] calculated over two consecutive points;
Col. [3]: Significance of the flux variation [$\sigma$];
Col. [4]: Maximum flux ratio measured over the whole light curve.
}
\centering
\begin{tabular}{lccc}
\hline
\hline
Band/Filter/Frequency & $\tau$ & $\sigma$ & $F_{\rm max}/F_{\rm min}$ \\
\hline   
$0.1-100$~GeV & $2.3\pm 0.5$ & 4.1 & 9.5\\
$0.3-10$~keV  & $6.7\pm 1.9$ & 4.9 & 4.9\\
UVW2          & $5.2\pm 2.8$ & 5.9 & 1.9\\
UVM2          & $<14$        & 3.9 & 1.9\\
UVW1          & $5.6\pm 5.0$ & 4.2 & 1.8\\
U             & $3.0\pm 1.1$ & 7.9 & 2.5\\
B             & $2.4\pm 0.8$ & 8.0 & 2.1\\
V             & $2.5\pm 2.1$ & 4.3 & 2.9\\
J             & $20\pm 8$    & 5.5 & 2.8\\
H             & $4.5\pm 1.5$ & 7.4 & 4.6\\
K             & $30\pm 3$    & 20.0  & 2.8\\
142~GHz       & $59\pm 11$   & 8.8 & 5.9\\
86~GHz        & $55\pm 14$   & 10.0  & 4.0\\
43~GHz        & $236\pm 46$  & 13.1 & 2.2\\
37~GHz        & $2.2\pm 1.8$ & 3.4  & 4.3\\
32~GHz        & $83\pm 23$   & 3.7  & 4.4\\
23~GHz        & $<244$       & 3.3  & 2.5\\
15~GHz        & $<6$         & 3.6  & 5.8\\
10.45~GHz     & $83\pm 6$    & 24.8 & 3.5\\
8.4~GHz       & $40\pm 22$   & 3.8  & 3.1\\
5~GHz      & $117\pm 42$  & 7.5  & 3.6\\
2.64~GHz      & $179\pm 20$  & 20.0 & 1.9\\
\hline
\hline 
\end{tabular}
\label{variability}
\end{table}

\begin{figure*}
\centering
\includegraphics[scale=0.45,clip, trim = 10 40 0 70]{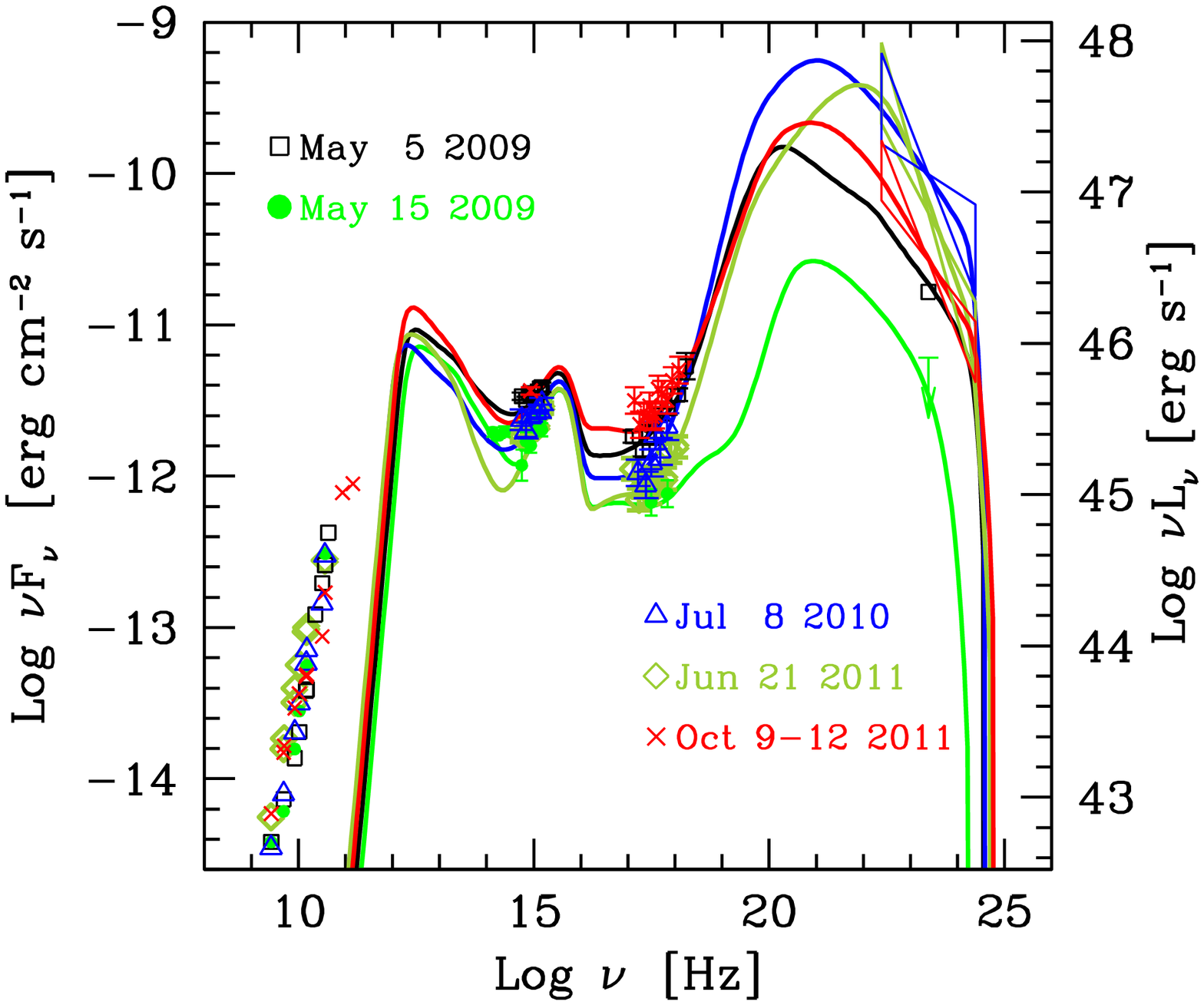}
\includegraphics[scale=0.45,clip, trim = 0 40 10 70]{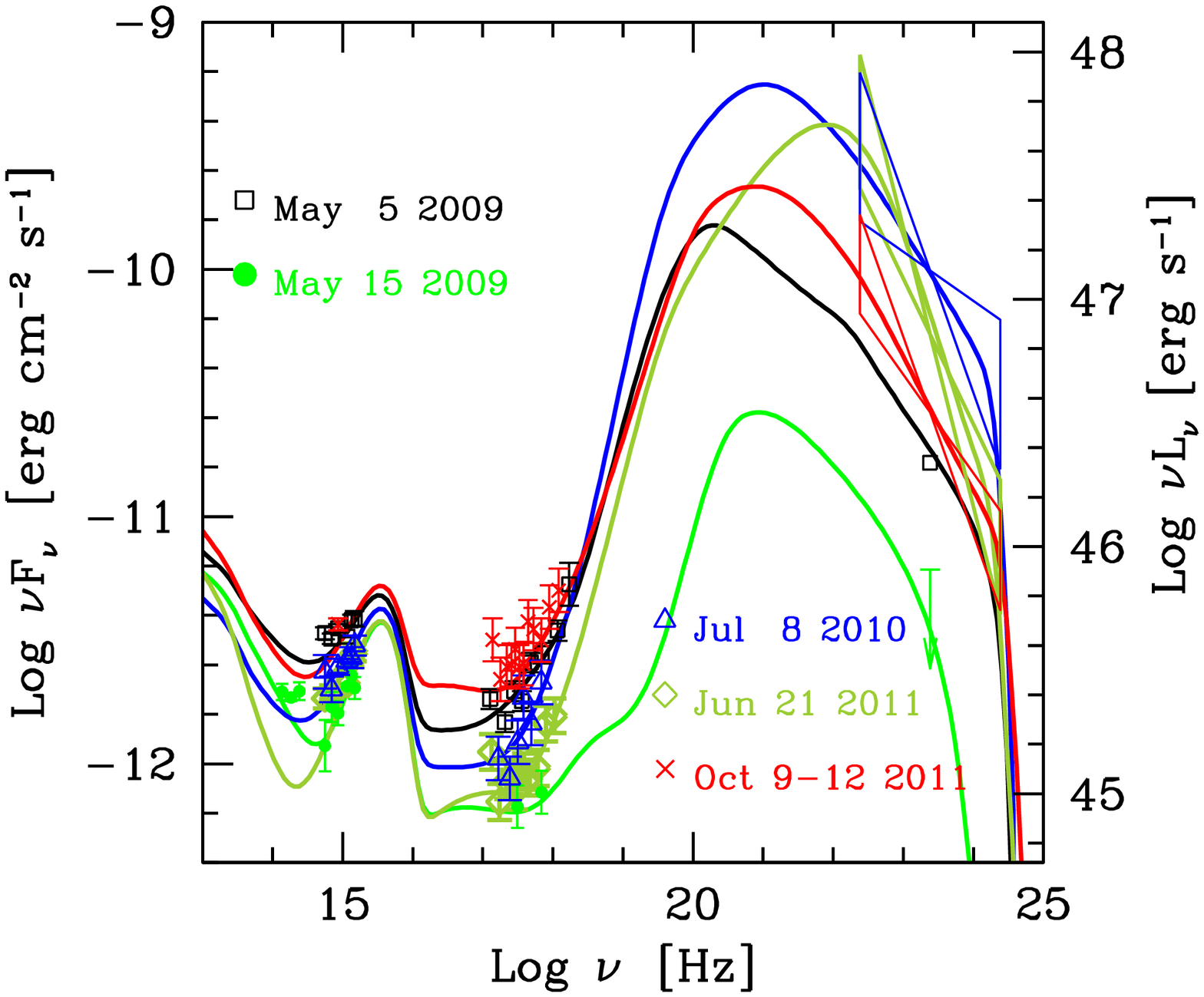}
\caption{({\it left panel}) Spectral Energy Distributions (SEDs) of PMN~J0948$+$0022 in some specific time periods. ({\it right panel}) Zoom of the infrared-to-$\gamma$-ray energy range. Each period is indicated with different symbols and colours, as displayed in the figures. Since we adopted an one-zone model, radio data were not fitted. The bow-ties at $\gamma$-rays indicate only the changes in the photon index, which are most relevant for the modelling. It is necessary to take into account that the flux normalisation has an error too, although not displayed. See the text for details.}
\label{Fig:SED}
\end{figure*}

\begin{table*}[!t]
\caption{Input parameters used to model the SED. 
Col. [1]: Date.
Col. [2]: distance of the blob from the black hole in units of $10^{15}$ cm; 
Col. [3]: $R_{\rm diss}$ in units of the Schwarzschild radius $R_{\rm S}$;
Col. [4]: power injected in the blob calculated in the comoving frame, in units of $10^{45}$ erg s$^{-1}$; 
Col. [5]: magnetic field in Gauss;
Col. [6]: bulk Lorentz factor;
Col. [7]:  random Lorentz factor of the injected electrons at the break of the distribution;
Col. [8]: maximum random Lorentz factor of the injected electrons;
Col. [9] and [10]: slopes of the injected electron distribution [$Q(\gamma)$] below 
and above $\gamma_{\rm b}$.
All models assume a black hole mass $M=1.5\times 10^8 M_\odot$ and an accretion disk luminosity
$L_{\rm d} =9\times 10^{45}$ erg s$^{-1}$, corresponding to $L_{\rm d}/L_{\rm Edd}=0.4$.
The X--ray corona has a luminosity $L_X=0.3 L_{\rm d}$.
The radius of the BLR is fixed to $R_{\rm BLR} = 3\times 10^{17}$ cm.
The spectral shape of the corona is assumed to be $\propto \nu^{-1} \exp(-h\nu/150~{\rm keV})$.
The viewing angle is $3^{\circ}$ for all models.
}
\centering
\begin{tabular}{lccccccccccc}
\hline
\hline
Date  &$R_{\rm diss}$ &$R_{\rm diss}/R_{\rm S}$   &$P^\prime_{\rm i}$  &$B$ &$\Gamma$  &$\gamma_{\rm b}$  &$\gamma_{\rm max}$ &$s_1$  &$s_2$ \\ 
~[1]       &[2] &[3] &[4] &[5] &[6] &[7] &[8] &[9] &[10]   \\
\hline   
2009 May 05       &67.5  &1500   &0.030    &4.4  &13  &10  &3000   &0    &2.65 \\     
2009 May 15       &49.5  &1100   &0.0055 &6.0  &11  &50  &1500 &0    &2.7 \\    
2010 Jul 08       &103.5 &2300   &0.044   &2.1  &16  &50  &3000   &0    &2.8 \\      
2011 Jun 21       &108.0    &2400   &0.025   &2.2  &16  &240 &4000   &0.5  &3.9 \\    
2011 Oct 09       &90.0    &2000   &0.032   &3.3  &12  &80  &5000   &1    &3.05 \\  
\hline
\hline 
\end{tabular}
\label{para}
\end{table*}

\section{Spectral Energy Distribution (SED) Modelling}
We selected some specific cases to represent the behaviour of PMN J0948$+$0022, and for which we had sufficient data. It is worth noting that the availability of X-ray observations played an important selection effect: indeed, as shown in Fig.~\ref{Fig:SED}, the peak of the synchrotron emission cannot be determined accurately, because of the synchrotron self-absorption. The X-ray emission is assumed to be due to the synchrotron self-Compton process; in this case, X-ray observations are required to estimate the magnetic field strength for modelling purposes. Therefore, even if the light curves (Fig.~\ref{Fig:curve}) indicate the presence of several interesting features, it was possible to properly model the corresponding SEDs only if an X-ray observation had been done within a few days.

The selected time periods are (see Fig.~\ref{Fig:curve}):

\begin{itemize}
\item 2009 May 5: early decrease of the optical-to-X-ray flux during the 2009 MW Campaign ({\it Fermi}/LAT data integrated over a time period of 5 days centred on the {\it Swift} observation; data at other wavelengths at the closest time);
\item 2009 May 15: lowest X-ray flux recorded during the 2009 MW Campaign and of the whole $\sim 3.5$ years ({\it Fermi}/LAT data integrated over a time period of 5 days centred on the {\it Swift} observation; data at other wavelengths at the closest time);
\item 2010 July 8: first $\gamma$-ray outburst ({\it Fermi}/LAT data integrated over one day; {\it Swift} observation on July 3; data at other wavelengths at the closest time);
\item 2011 June 21: second $\gamma$-ray outburst ({\it Fermi}/LAT data integrated over one day; {\it Swift} observation on June 14; data at other wavelengths at the closest time);
\item 2011 October 9-12: maximum optical-to-X-ray flux ({\it Fermi}/LAT data integrated over a time period of 5 days centred on October 11; two {\it Swift} observations on October 9 and 12 integrated; data at other wavelengths at the closest time).
\end{itemize}

The corresponding SEDs and models are displayed in Fig.~\ref{Fig:SED}. It is worth noting that the data are not strictly simultaneous. As written above, the zero-order rule was to use the {\it Swift} observation as the main driver, but obviously there were major events (e.g. the $\gamma$-ray outbursts) that cannot be ignored. When possible, we used the closest (within one week) {\it Swift} pointing, which were on 2010 July 3 for the 2010 outburst (i.e. 5 days before) and on 2011 June 14 for the 2011 outburst (i.e. 7 days before). In 2011 there were other outbursts at $\gamma$-rays (2011 July 28, day 1091; 2011 September 25, day 1150), but no timely {\it Swift} observations were available. 

We adopted the same model used in past MW campaigns (LAT Coll. 2009c, Foschini et al. 2011), which is that developed by Ghisellini \& Tavecchio (2009). It is a one-zone model, which calculates the emitted radiation from a population of relativistic electrons through synchrotron self-Compton and external Compton (EC) processes. In the latter, the seed photons are considered from different sources (accretion disk, broad-line region, infrared torus, etc.). We refer to Ghisellini \& Tavecchio (2009) for more details on the model and to LAT Coll. (2009c) and Foschini et al. (2011) for the specific application to the 2009 and 2010 MW campaigns on PMN~J0948$+$0022, respectively, with specific reference to the possible changes in the model parameters as a function of the data. 

The input parameters (see Table~\ref{para}) were adjusted to fit the data and the model output consists of the calculated jet power divided into its basic components (radiative and kinetic, see Table~\ref{powers}). Given the large number of parameters, we fixed some of them by taking into account the available measured quantities. The mass of the central black hole is $M\sim 1.5\times 10^{8} M_{\odot}$, as derived from the fit to the accretion disk emission (LAT Coll. 2009b), which is consistent with the black hole mass estimate by Zhou et al. (2003) by using the classical virial method. The accretion disk luminosity is measured as $L_{\rm d} \sim 9\times 10^{45}$~erg~s$^{-1}$, corresponding to $L_{\rm d}/L_{\rm Edd} \sim 0.4$. The value has been measured during the period of minimum jet flux and fixed on short time scales, although it can change over the years. 

The jet viewing angle is fixed to $\sim 3^{\circ}$ for all SEDs. In earlier work, we adopted a value of $6^{\circ}$ (LAT Coll. 2009b,c), but modelling of the 2010 July outburst required a smaller viewing angle to explain the observed $\gamma$-ray isotropic luminosity (Foschini et al. 2011). The smaller jet angle generates slightly different parameter values compared to LAT Coll. (2009c), but does not significantly change the interpretation. 

In addition, fixing the above parameters has some implications on the other parameters. For example, as known from the reverberation mapping, the radius of the broad-line region (BLR) is a function of the accretion disk luminosity (see Eq.~2 in Ghisellini \& Tavecchio 2009). The values of the fitted parameters are presented in Table~\ref{para} and the calculated output jet powers are shown in Table~\ref{powers}.

\begin{table}[!t] 
\caption{Calculated jet power in the form of radiation, Poynting flux, bulk motion of electrons and protons (assuming one cold proton per emitting electron). The logarithms are calculated on the powers expressed in erg~s$^{-1}$.}
\centering
\begin{tabular}{lcccc}
\hline
\hline
Date   &$\log P_{\rm r}$ &$\log P_{\rm B}$ &$\log P_{\rm e}$ &$\log P_{\rm p}$  \\
\hline  
2009 May 05     &45.31 &44.75 &45.23 &47.49     \\
2009 May 15     &44.55 &44.59 &44.33 &46.17     \\         
2010 Jul 08     &45.97 &44.67 &45.43 &47.61     \\
2011 Jun 21     &45.76 &44.47 &44.98 &47.10     \\
2011 Oct 09     &45.47 &44.68 &45.20 &47.46     \\       
\hline
\hline 
\end{tabular}
\label{powers}
\end{table}

The modelled SEDs can be interpreted by looking at the most relevant changes in the parameters. The drop in $\gamma$-ray emission in early 2009 May can be basically explained with a decrease in the injected power (from $3\times 10^{43}$~erg~s$^{-1}$ to $5.5\times 10^{42}$~erg~s$^{-1}$), an increase of the magnetic field (from 4.4 to 6.0 G), and a decrease of the bulk Lorentz factor (from 13 to 11). The slope of the electron distribution remained almost constant. On the other hand, the 2010 July outburst requires a high injected power ($4.4\times 10^{43}$~erg~s$^{-1}$) and also a higher bulk Lorentz factor ($\Gamma=16$), together with a smaller magnetic field (2.1~G). The 2011 June outburst, although with similar magnetic field and Lorentz factor (2.2~G and 16, respectively), requires less injected power ($2.5\times 10^{43}$~erg~s$^{-1}$), but a softer electron distribution. Indeed, as noted in Sect.~2.1, the LAT spectrum showed in 2011 softening in the slope. 

A comparison of the values obtained by the SEDs modeling of PMN~J0948$+$0022 (Table~\ref{para} and \ref{powers}) with those of a large sample of bright blazars (Ghisellini et al. 2010) shows that the jet powers are in the range of flat-spectrum radio quasars, although there are some differences in the input parameters. The average mass  and accretion luminosity of quasars in the sample of Ghisellini et al. (2010) are $\sim 10^{9}M_{\odot}$ and $\sim 0.1L_{\rm Edd}$, respectively, to be compared with the values of PMN~J0948$+$0022 ($M \sim 10^{8}M_{\odot}$, $L_{\rm d}\sim 0.4L_{\rm Edd}$). However, it is possible to find quasars with similar characteristics, like PKS~0426$-$380 ($z=1.112$), which has $M\sim 4\times 10^{8}M_{\odot}$ and $L_{\rm d}\sim 0.6L_{\rm Edd}$. The input and output parameters of the model (see Table 4 and 5 in Ghisellini et al. 2010) are very similar to those presented here.

\section{Conclusions}
The most likely physical interpretation for the observed data is that the NLS1 galaxy PMN~J0948$+$0022 hosts a powerful relativistic jet directed, within a few degrees, toward the Earth, confirming previous works. Zhou et al. (2003) and Doi et al. (2006), having found an inverted radio spectrum with high brightness temperature in excess of $10^{13}$~K, suggested that the NLS1 galaxy should host a relativistic jet. However, there are some caveats in inferring the presence of a relativistic jet from the high-brightness temperature, as outlined for example by Tsang \& Kirk (2007) and Singal (2009). A halo of doubt remained, particularly given the fact that it was quite anomalous and strange that a NLS1 type AGN, which is generally radio-quiet, hosted a relativistic jet.

The detection of variable high-energy $\gamma$-ray emission from PMN~J0948$+$0022 with {\it Fermi}/LAT (LAT Coll. 2009b, 2010a) provided the needed breakthrough. Specifically, the 2009 MW campaign, which found coordinated MW variability, confirmed beyond any doubt the association of the high-energy $\gamma$-ray source with the NLS1 (LAT Coll. 2009c). In addition, the observation of optical ($V$ filter) polarisation at 19\% level (Ikejiri et al. 2011) and optical ($B$ and $R$ filters) intraday variability (Liu et al. 2010), provided useful and important complementary information to strengthen the relativistic jet scenario. 

The reanalysis of the data presented in this work confirm and extend the early findings reported by LAT Coll. (2009b,c, 2010a). We have selected five epochs to represent the source activity during these three and a half years. The most relevant parameters driving the modelling of the SEDs in 2009 and 2010 are the injected power, the bulk Lorentz factor, and the magnetic field. The 2011 SEDs are instead characterised by a change (softening) in the electron distribution, with a decreasing contribution of high-energy leptons. The analysis of the SEDs modelling, together with the study of the variability, indicates that the jet hosted by the NLS1 PMN~J0948$+$0022 is like those hosted by the flat-spectrum radio quasar subclass of blazars. This is another point favouring the universality of the jet phenomenon.

In the years 2009 and 2010, the jet emission followed what is  expected from the canonical model of relativistic jets (Blandford \& K\"onigl 1979, Heinz \& Sunyaev 2003), where the peak of radio emission follows the $\gamma$-rays after a few months. The year 2011 was instead characterised by a prolonged activity, with some peculiar events. Specifically, there was an optical/X-ray flare, which has no corresponding activity at other wavelengths. Such ``orphan flares''\footnote{Although this term has been created by H. Krawczynski when speaking about a VHE flare of 1ES 1959$+$650 with no counterpart at other wavelengths (Krawczynski et al. 2004), its meaning can be extended to any flare at any frequency with no counterparts at other frequencies.} have been observed also in other blazars (e.g. Krawczynski et al. 2004, LAT Coll. 2010d, VERITAS Coll. 2011, Marscher 2012, Marscher et al. 2012). Some attempts to explain this behaviour have been suggested (e.g. Marscher 2012), but none seems to meet with the general consensus. In the present case, the sparse and irregular sampling at NIR/optical/UV/X-rays prevents the testing of specific hypotheses. The observed variability of a few days suggests the need of a denser sampling rate, perhaps a MW campaign limited to a short period, but with day-scale sampling.

\section*{Acknowledgments}
We acknowledge the internal referee of the {\it Fermi}/LAT Collaboration, F. D'Ammando, for useful comments.

This research is partly based on observations with the 100-m telescope of the MPIfR (Max-Planck-Institut f\"ur Radioastronomie) at Effelsberg and with the IRAM 30-m telescope. IRAM is supported by INSU/CNRS (France), MPG (Germany) and IGN (Spain). I. Nestoras is funded by the International Max Planck Research School (IMPRS) for Astronomy and Astrophysics at the Universities of Bonn and Cologne.

The Mets\"ahovi team acknowledges the support from the Academy of Finland to our observing projects (numbers 212656, 210338, 121148, and others)

The OVRO 40-m monitoring program is supported in part by NASA grants NNX08AW31G and NNX11A043G, and NSF grants AST-0808050 and AST-1109911.

The National Radio Astronomy Observatory is a facility of the National Science Foundation operated under cooperative agreement by Associated Universities, Inc. This research has made use of data from the MOJAVE database that is maintained by the MOJAVE team (Lister et al. 2009). The MOJAVE project is supported NASA-Fermi grant NNX08AV67G. This work made use of the Swinburne University of Technology software correlator, developed as part of the Australian Major National Research Facilities Programme and operated under licence. YYK is partly supported by the Russian Foundation for Basic Research (project 11-02-00368), the basic research program ``Active processes in galactic and extragalactic objects'' of the Physical Sciences Division of the Russian Academy of Sciences and the Dynasty Foundation. 

This work is partially supported by Grant-in-Aid for Scientific Researches, KAKENHI 24540240 (MK) and 24340042 (AD) from Japan Society for the Promotion of Science (JSPS).

The \textit{Fermi} LAT Collaboration acknowledges generous ongoing support from a number of agencies and institutes that have supported both the development and the operation of the LAT as well as scientific data analysis. These include the National Aeronautics and Space Administration and the Department of Energy in the United States, the Commissariat \`a l'Energie Atomique and the Centre National de la Recherche Scientifique / Institut National de Physique Nucl\'eaire et de Physique des Particules in France, the Agenzia Spaziale Italiana and the Istituto Nazionale di Fisica Nucleare in Italy, the Ministry of Education, Culture, Sports, Science and Technology (MEXT), High Energy Accelerator Research Organization (KEK) and Japan Aerospace Exploration Agency (JAXA) in Japan, and the K.~A.~Wallenberg Foundation, the Swedish Research Council and the Swedish National Space Board in Sweden. Additional support for science analysis during the operations phase is gratefully acknowledged from the Istituto Nazionale di Astrofisica in Italy and the Centre National d'\'Etudes Spatiales in France.

Swift at PSU is supported by NASA contract NAS5-00136.

SK would like to thank the Aspen Center for Physics for their hospitality. The Aspen Center for Physics is supported by NSF Grant \#1066293. 

This work has been partially supported by ASI-INAF Grant I/009/10/0.

This research has made use of data obtained from the High Energy Astrophysics Science Archive Research Center (HEASARC), provided by NASA's Goddard Space Flight Center. 

Funding for the SDSS and SDSS-II has been provided by the Alfred P. Sloan Foundation, the Participating Institutions, the National Science Foundation, the U.S. Department of Energy, the National Aeronautics and Space Administration, the Japanese Monbukagakusho, the Max Planck Society, and the Higher Education Funding Council for England. The SDSS Web Site is http://www.sdss.org/. The SDSS is managed by the Astrophysical Research Consortium for the Participating Institutions. The Participating Institutions are the American Museum of Natural History, Astrophysical Institute Potsdam, University of Basel, University of Cambridge, Case Western Reserve University, University of Chicago, Drexel University, Fermilab, the Institute for Advanced Study, the Japan Participation Group, Johns Hopkins University, the Joint Institute for Nuclear Astrophysics, the Kavli Institute for Particle Astrophysics and Cosmology, the Korean Scientist Group, the Chinese Academy of Sciences (LAMOST), Los Alamos National Laboratory, the Max-Planck-Institute for Astronomy (MPIA), the Max-Planck-Institute for Astrophysics (MPA), New Mexico State University, Ohio State University, University of Pittsburgh, University of Portsmouth, Princeton University, the United States Naval Observatory, and the University of Washington.

\Online

\begin{appendix}
\section{Light curves}
Light curves of all the data available and studied in the present work are displayed in the following. 

\begin{figure*}
\centering
\includegraphics[angle=270,scale=0.3]{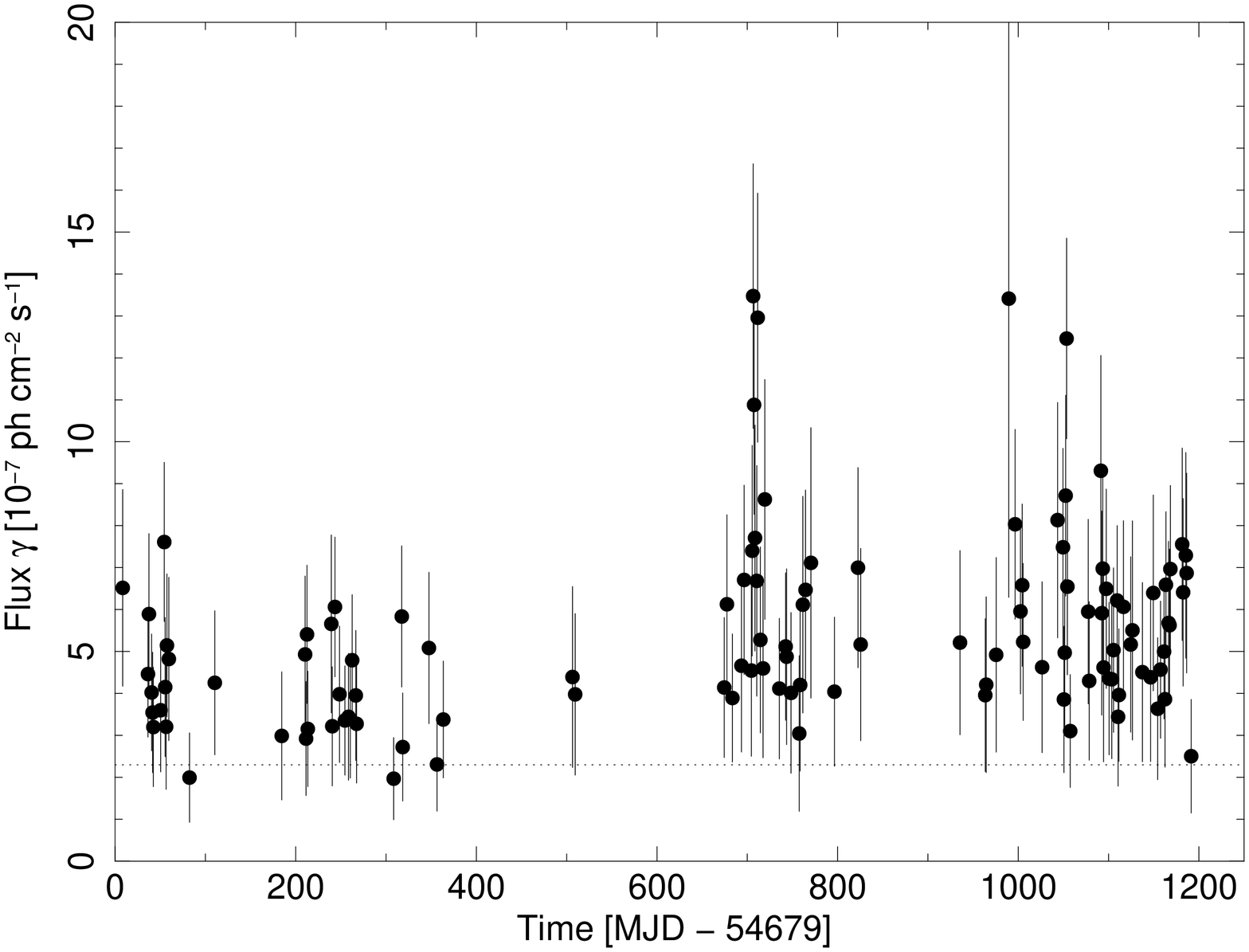}
\includegraphics[angle=270,scale=0.3]{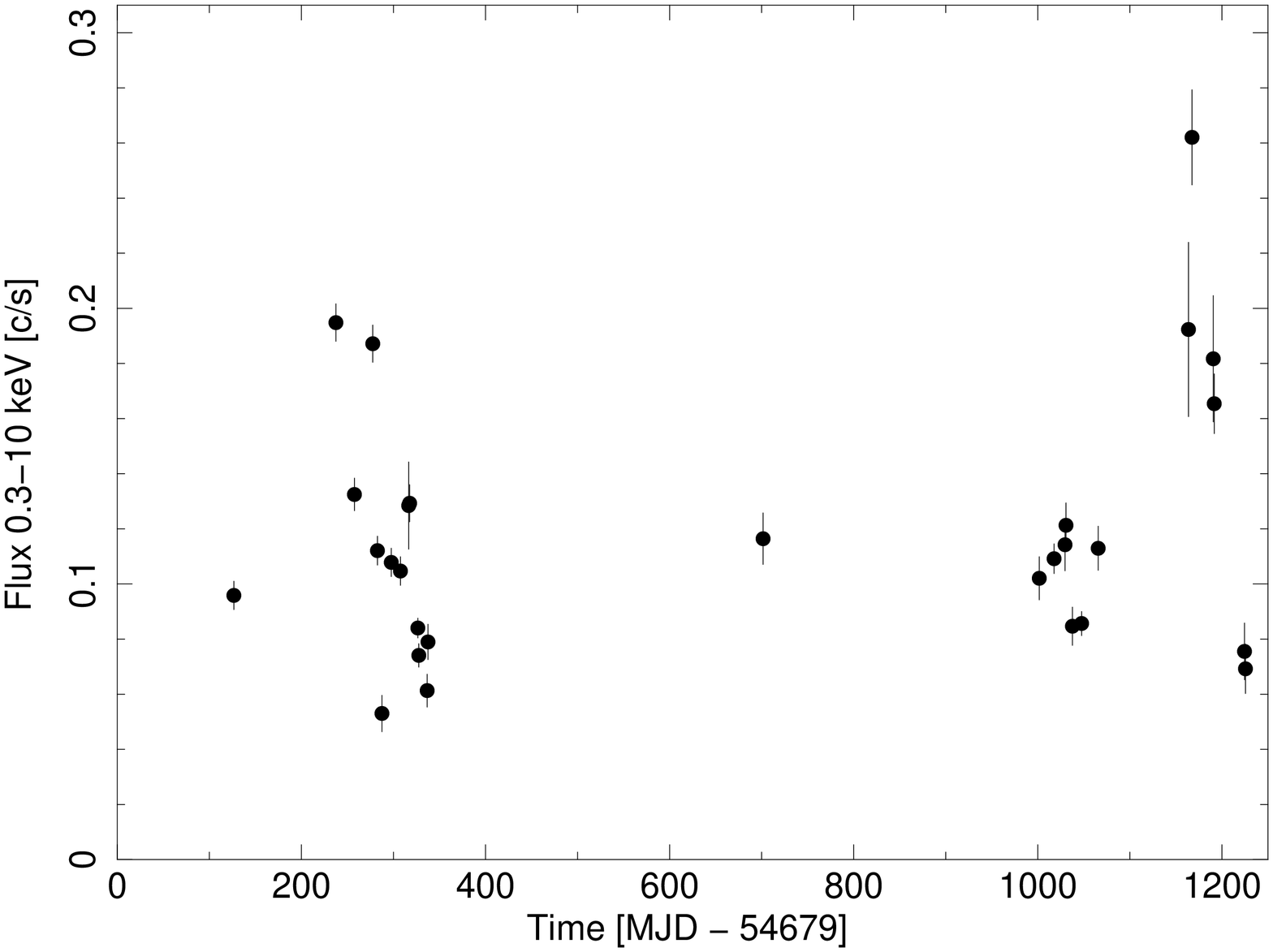}\\
\includegraphics[angle=270,scale=0.3]{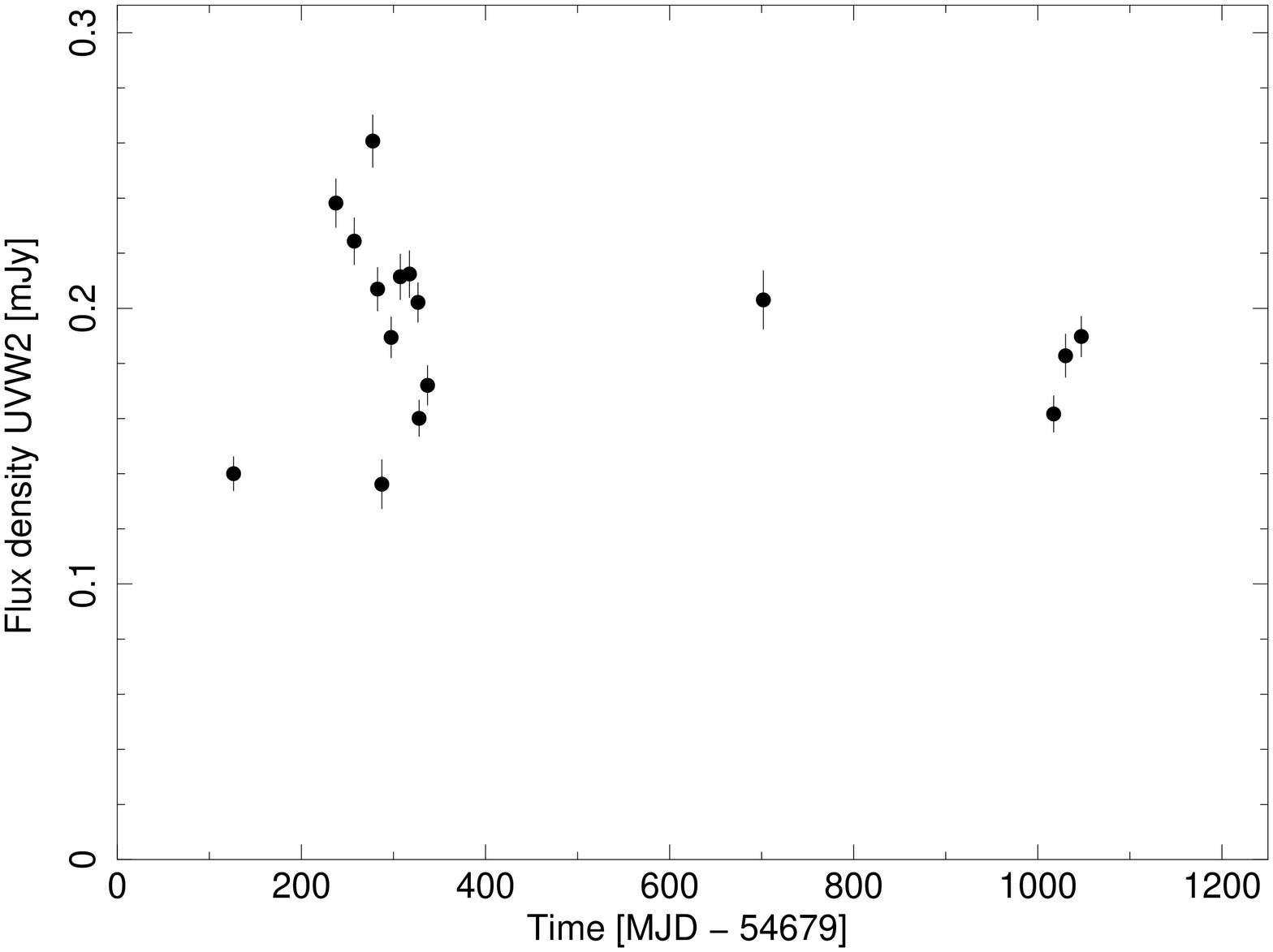}
\includegraphics[angle=270,scale=0.3]{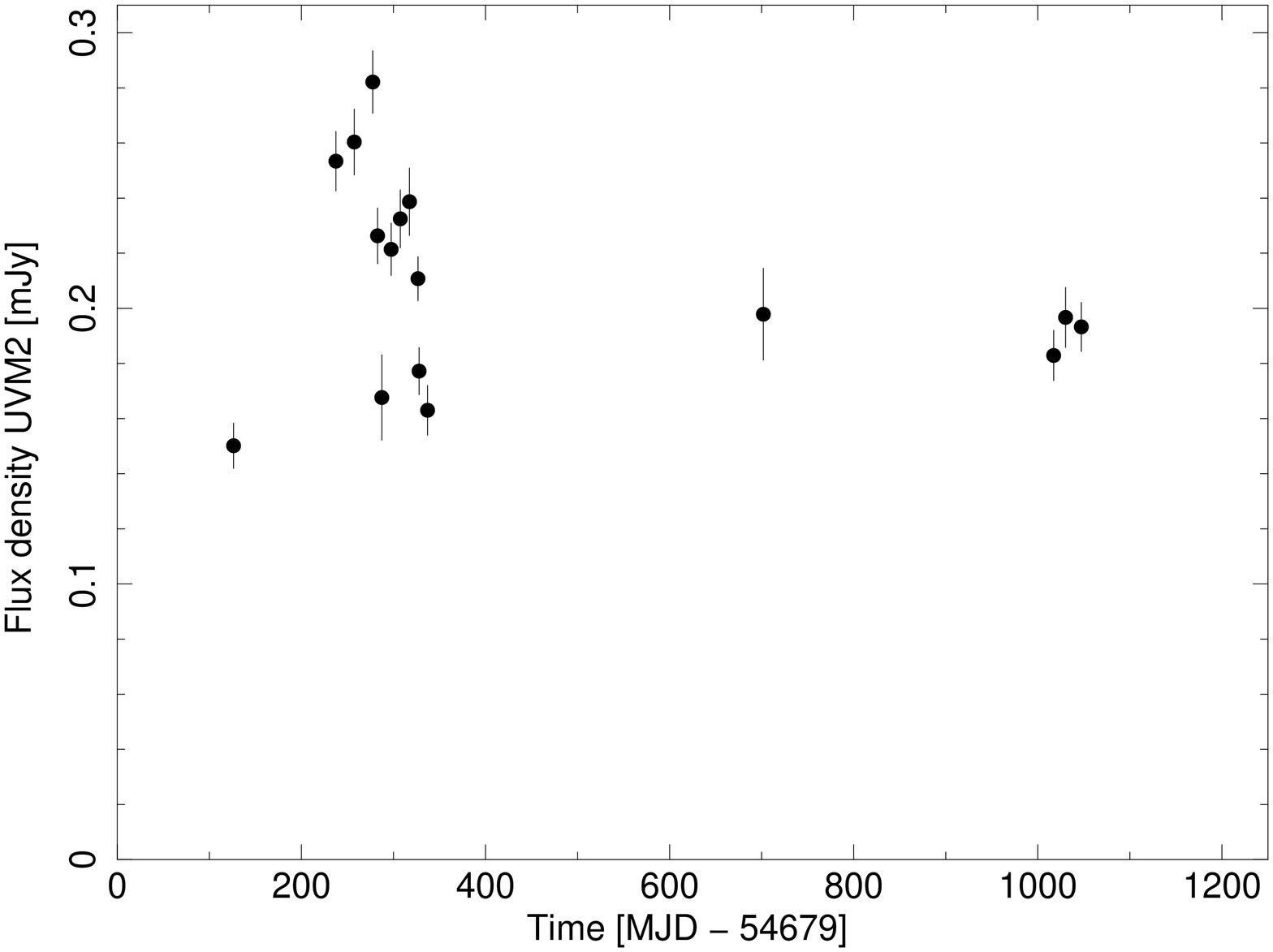}\\
\includegraphics[angle=270,scale=0.3]{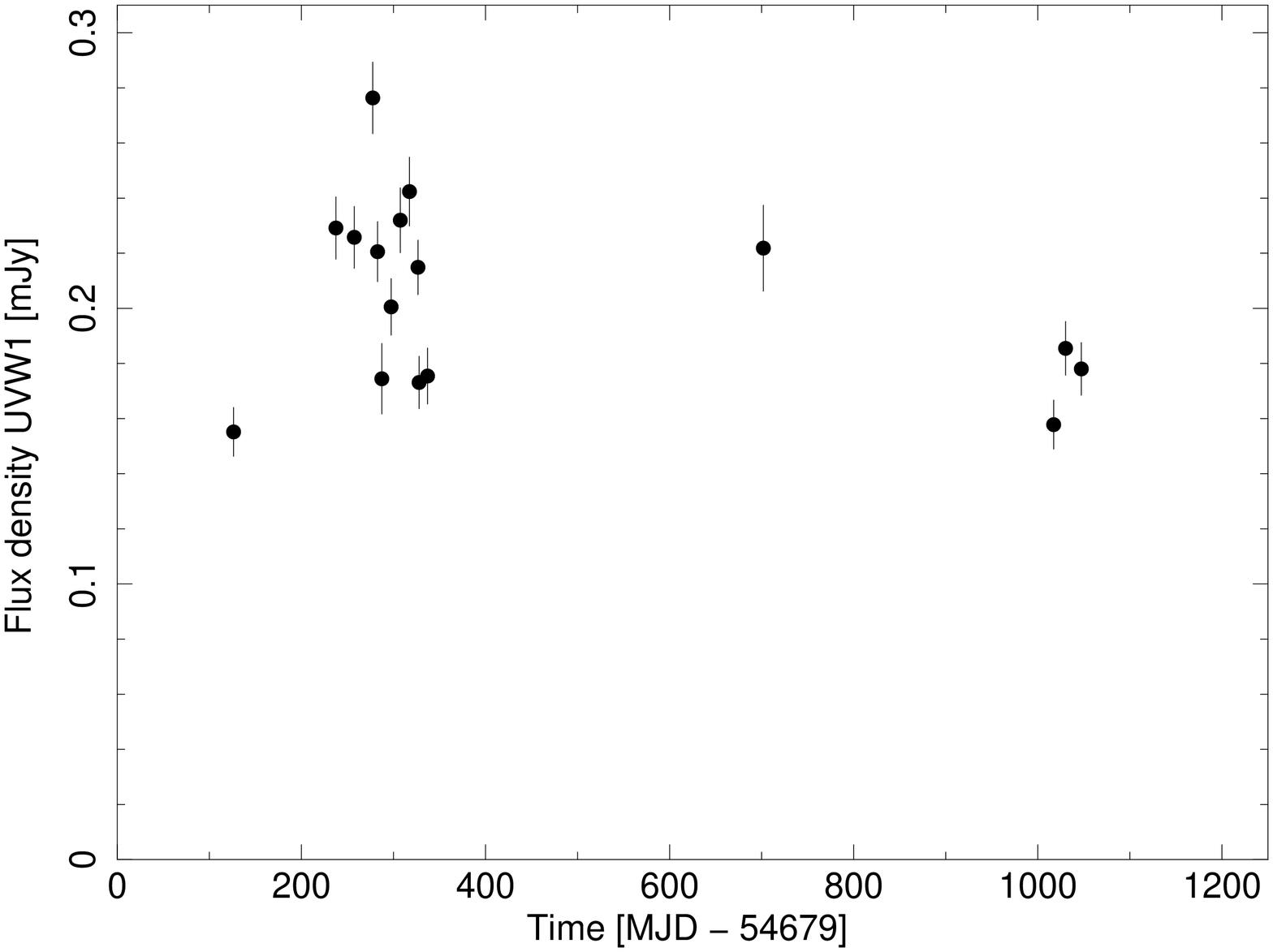}
\includegraphics[angle=270,scale=0.3]{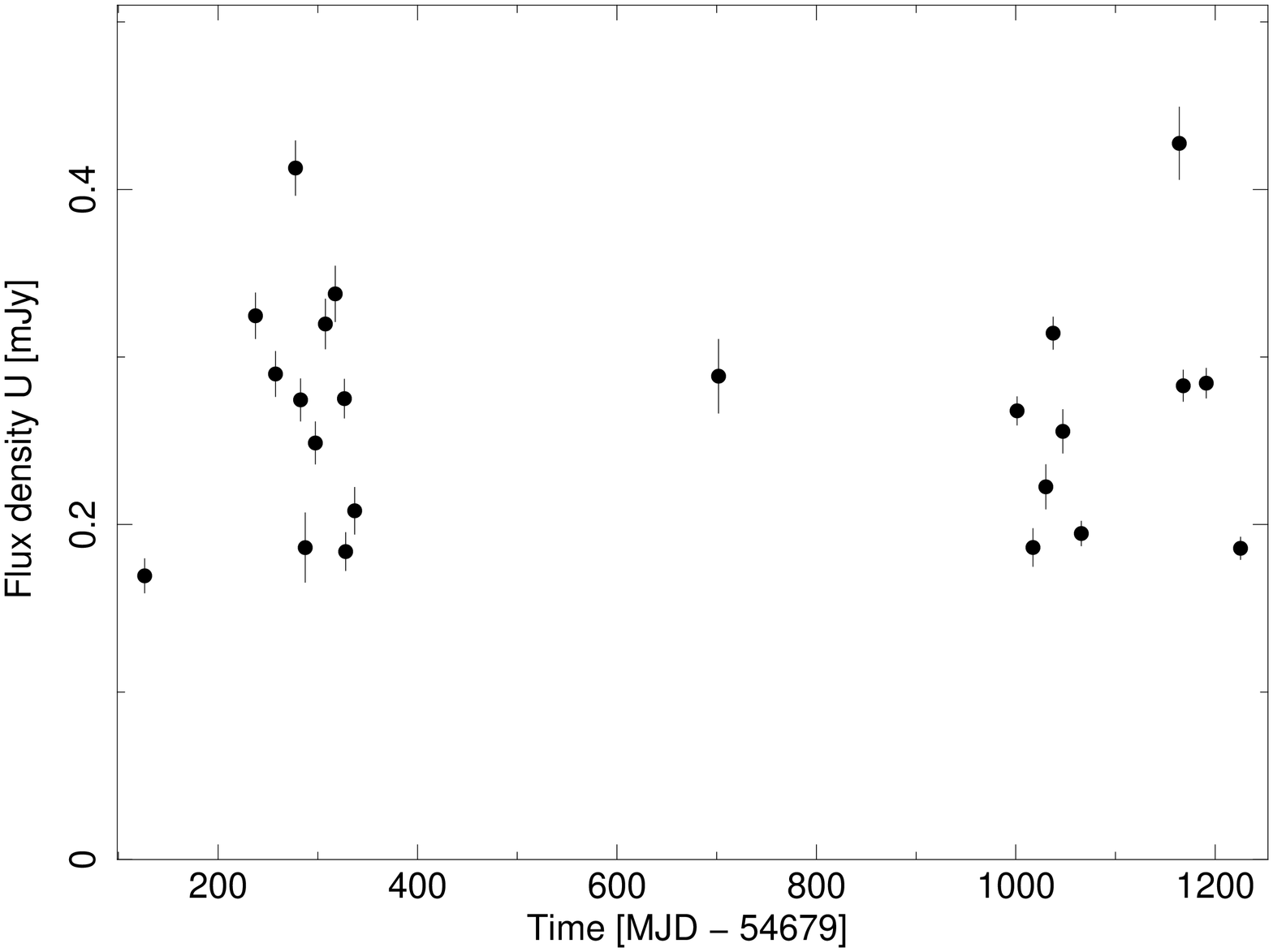}\\
\caption{Light curves at various frequencies. From top left to bottom right panel: $\gamma$-ray $0.1-300$~GeV from {\it Fermi}/LAT, with 1 day time bin [$10^{-7}$~ph~cm$^{-2}$~s$^{-1}$] (the horizontal dotted line correspond to a $TS=4$ average upper limit of $2.3\times 10^{-7}$~ph~cm$^{-2}$~s$^{-1}$ for a one-day exposure); X-ray $0.3-10$~keV from {\it Swift}/XRT [c~s$^{-1}$]; UVW2 from {\it Swift}/UVOT [mJy]; UVM2 from {\it Swift}/UVOT [mJy]; UVW1 from {\it Swift}/UVOT [mJy]; U from {\it Swift}/UVOT [mJy]. Time starts on 2008 August 1 00:00 UTC (MJD 54679).}
\label{Fig:CURVE1}
\end{figure*}

\begin{figure*}
\centering
\includegraphics[angle=270,scale=0.3]{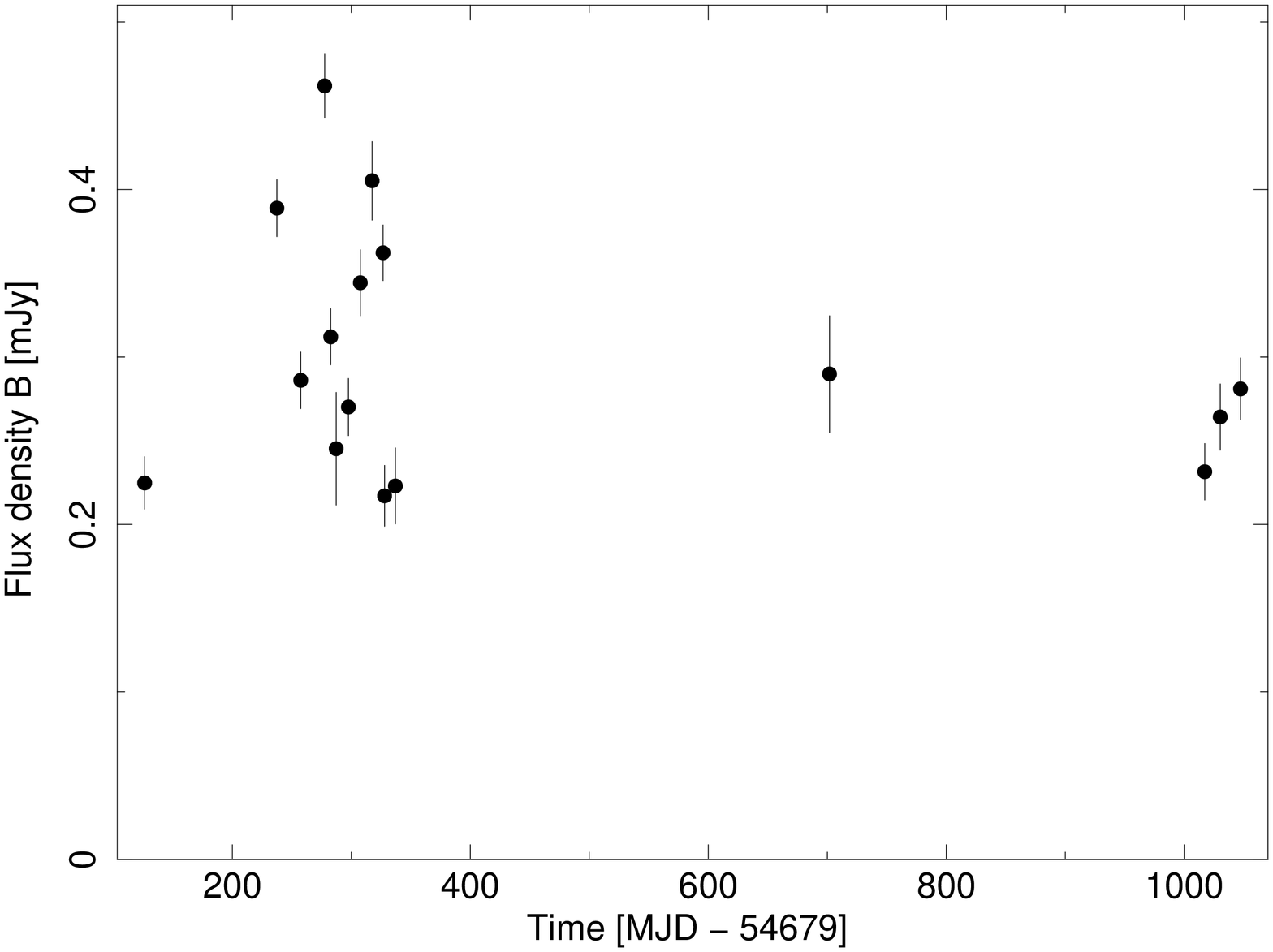}
\includegraphics[angle=270,scale=0.3]{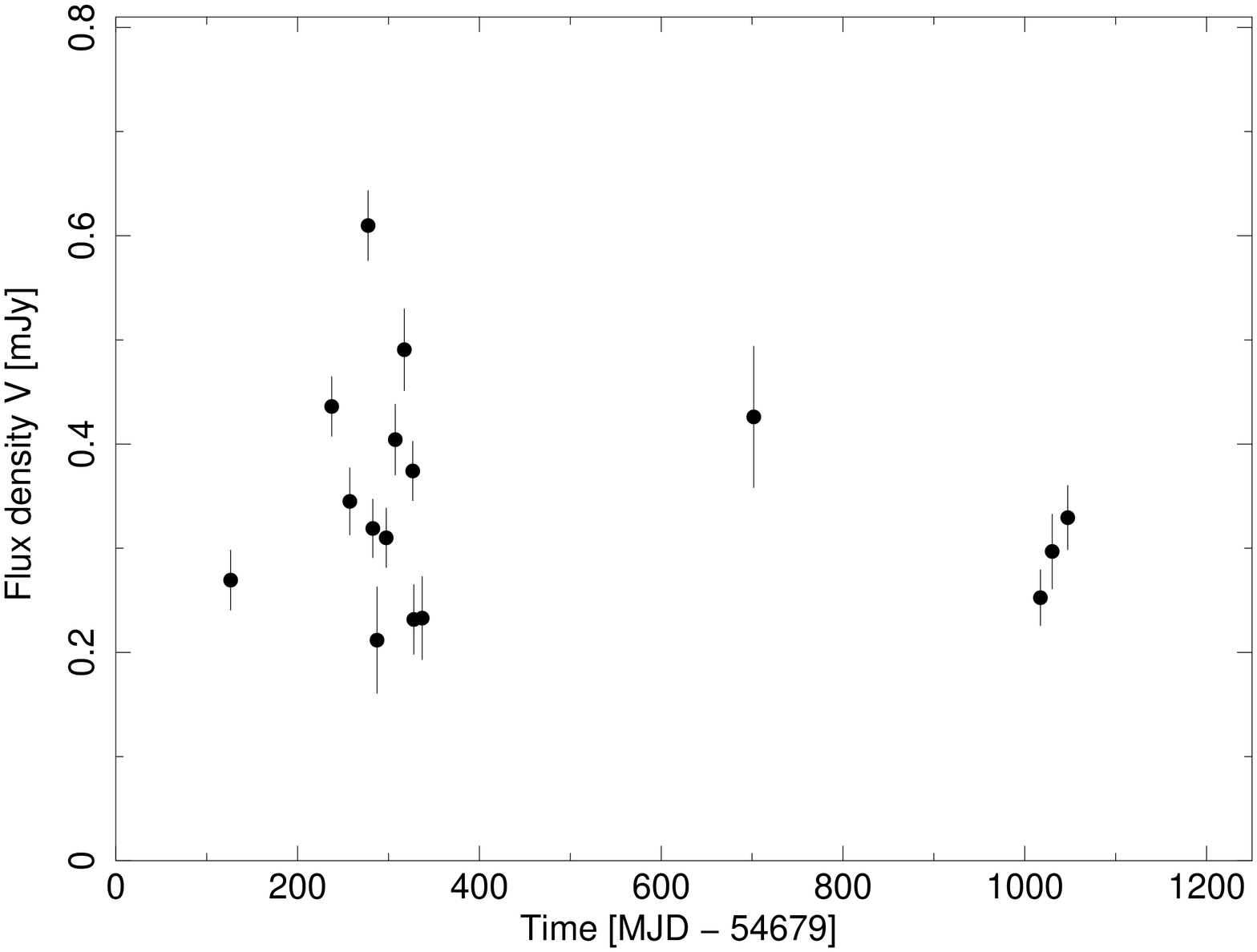}\\
\includegraphics[angle=270,scale=0.3]{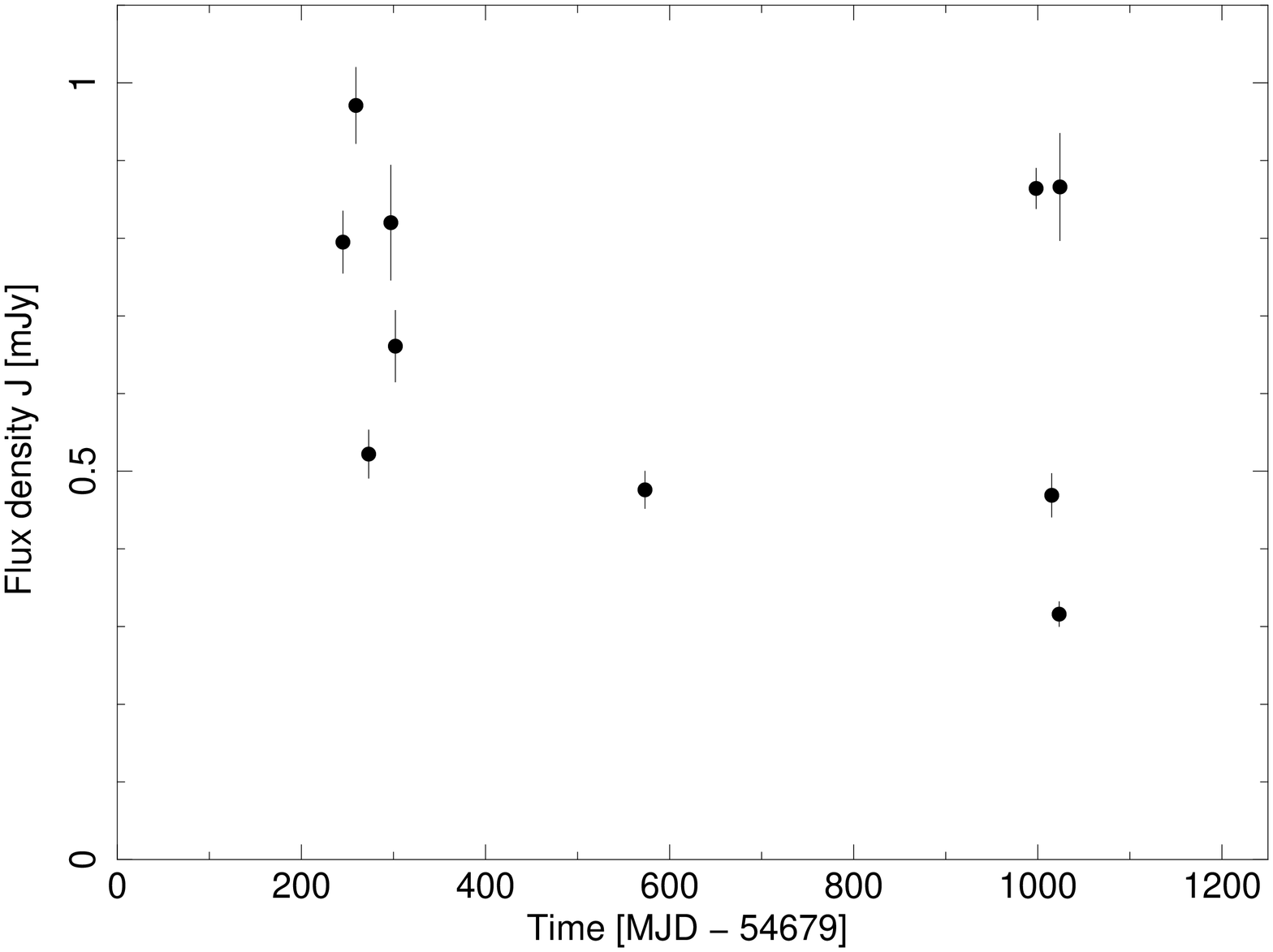}
\includegraphics[angle=270,scale=0.3]{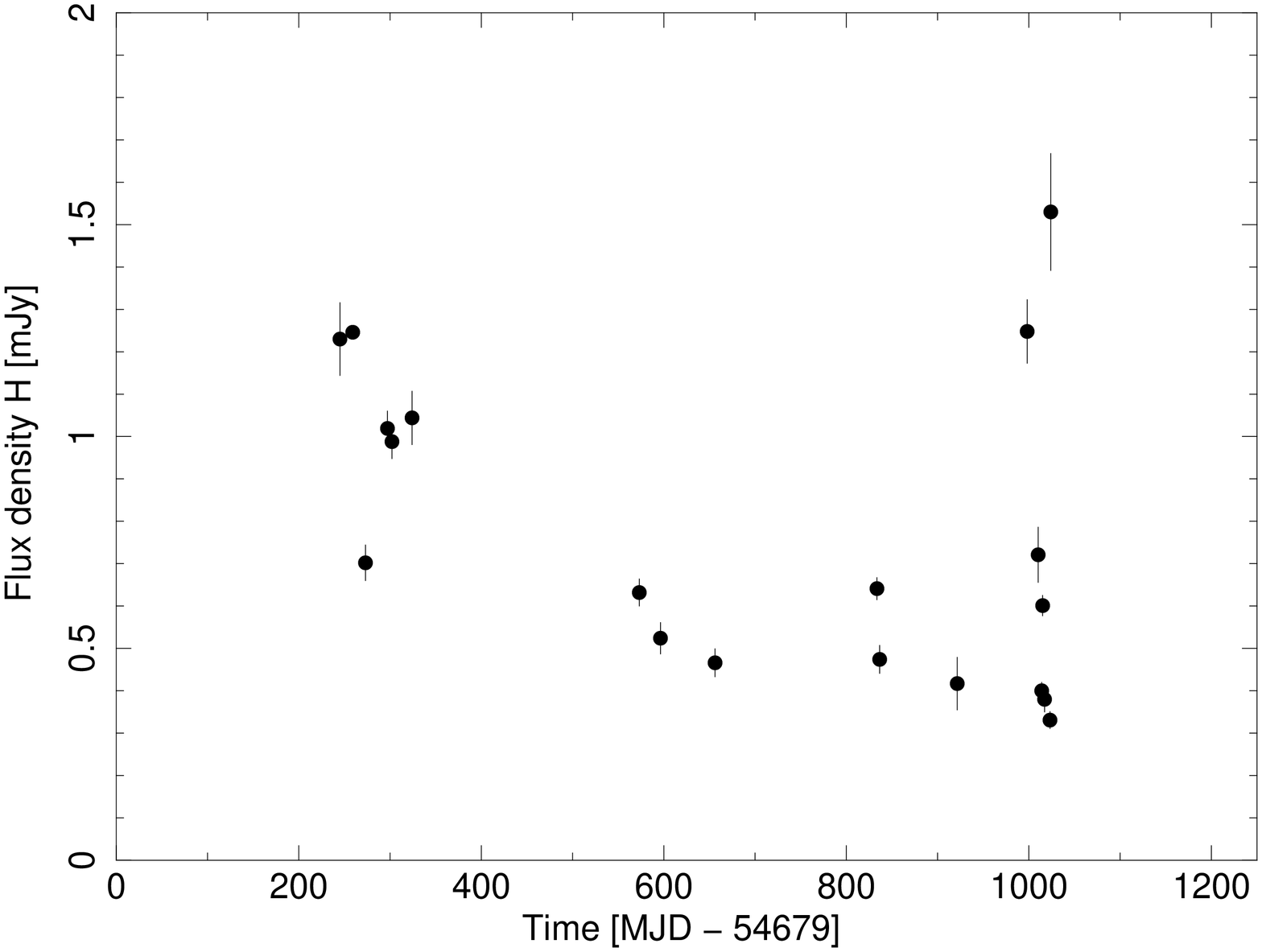}\\
\includegraphics[angle=270,scale=0.3]{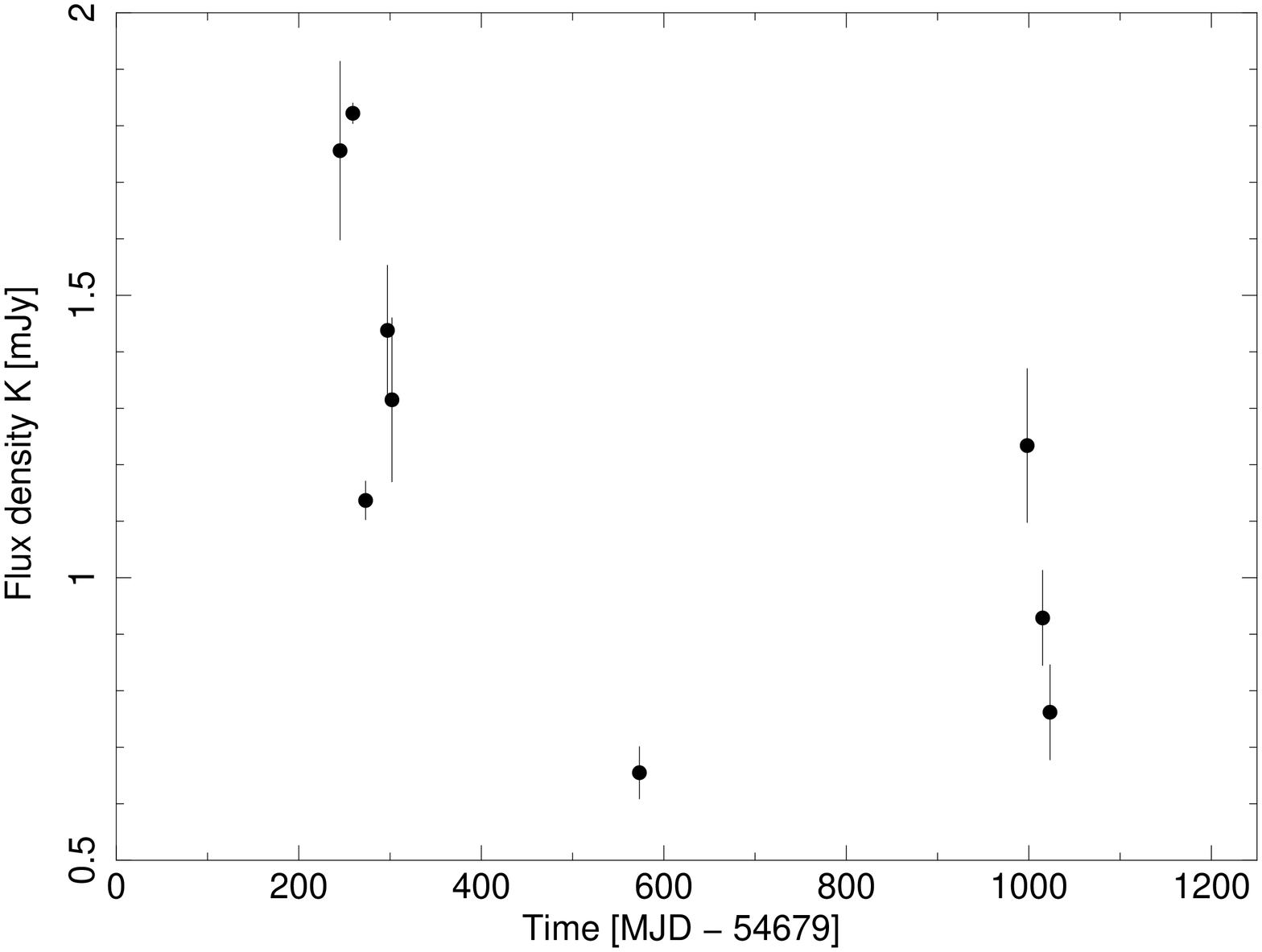}
\includegraphics[angle=270,scale=0.3]{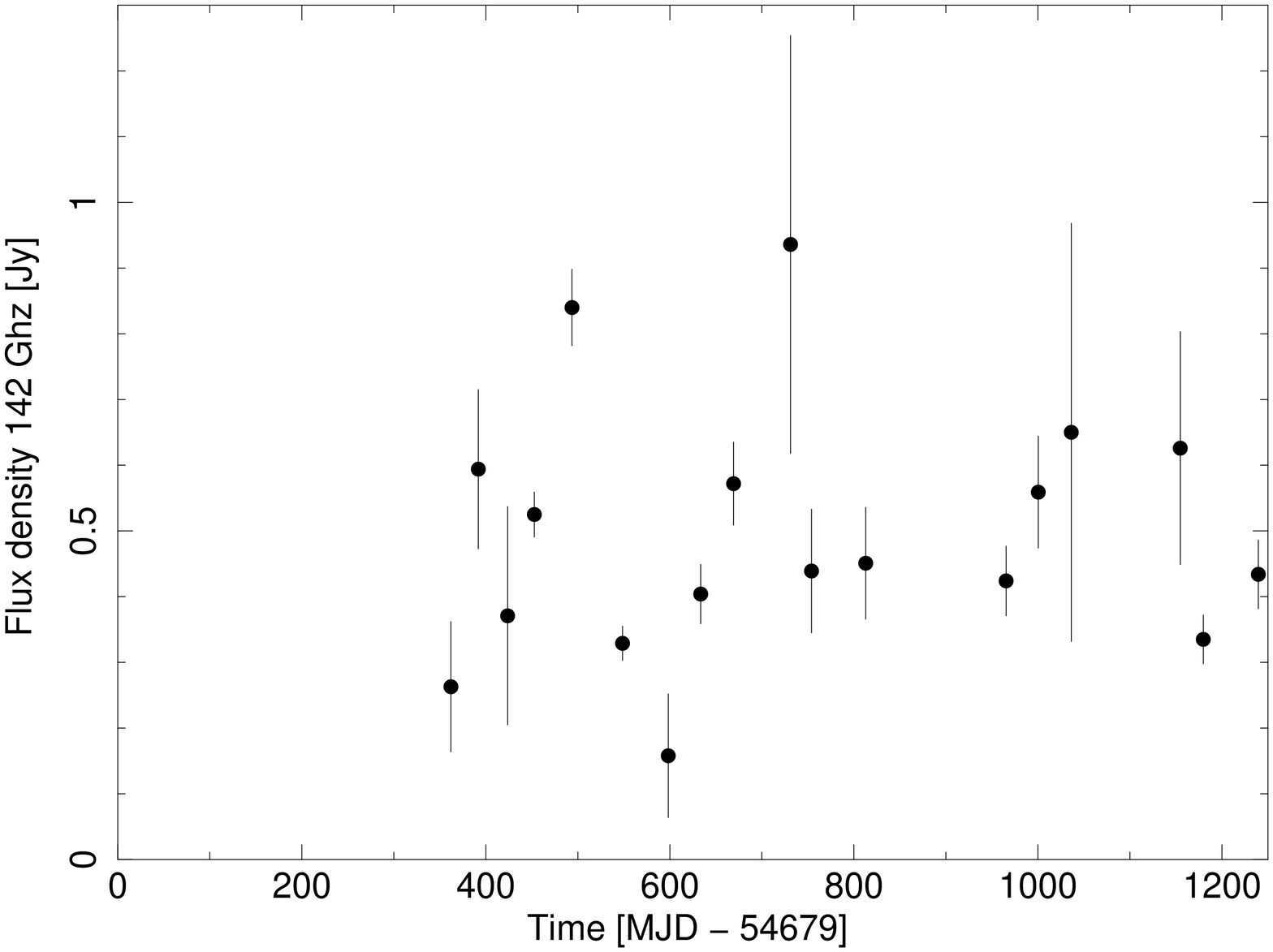}\\
\includegraphics[angle=270,scale=0.3]{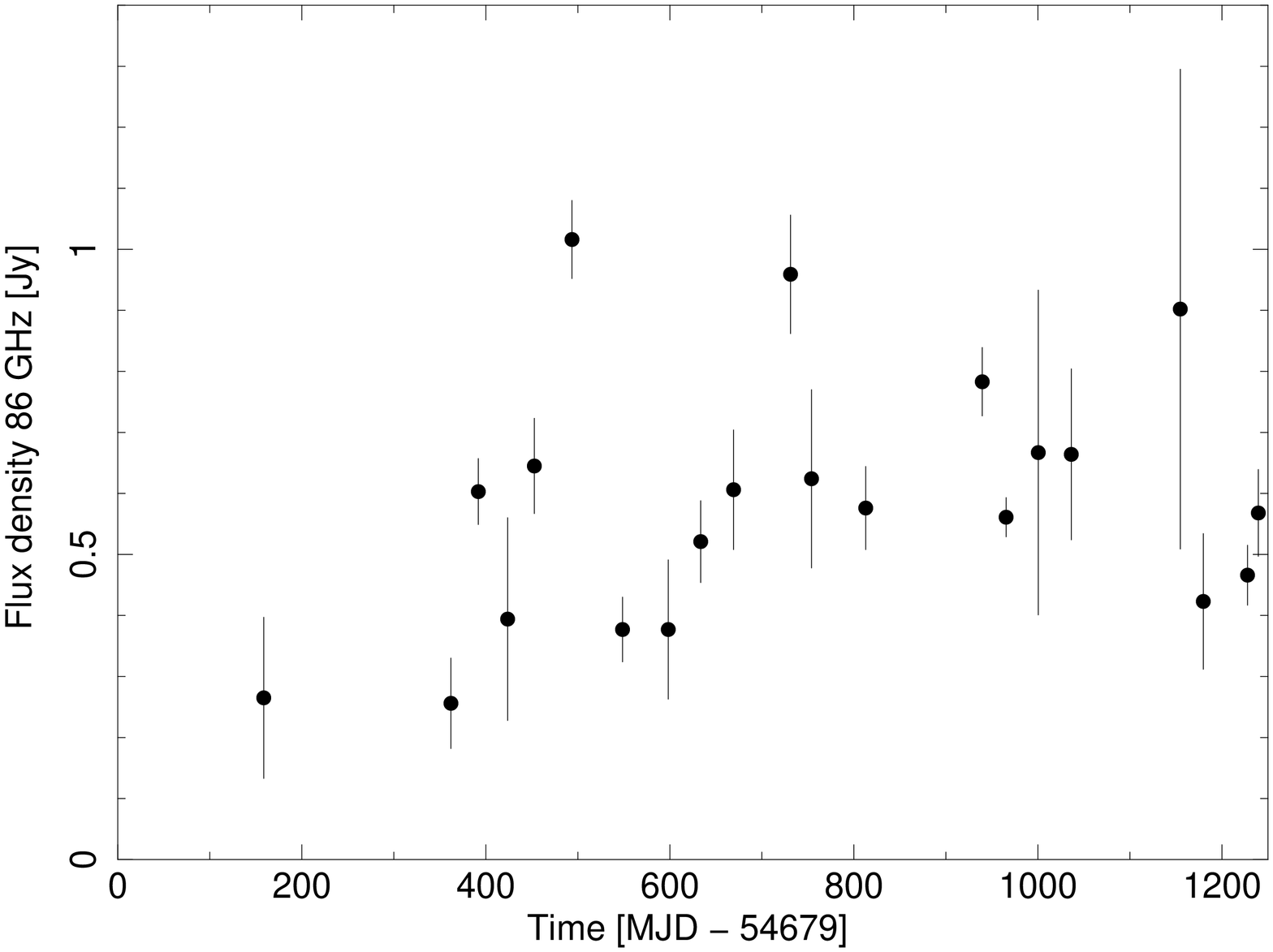}
\includegraphics[angle=270,scale=0.3]{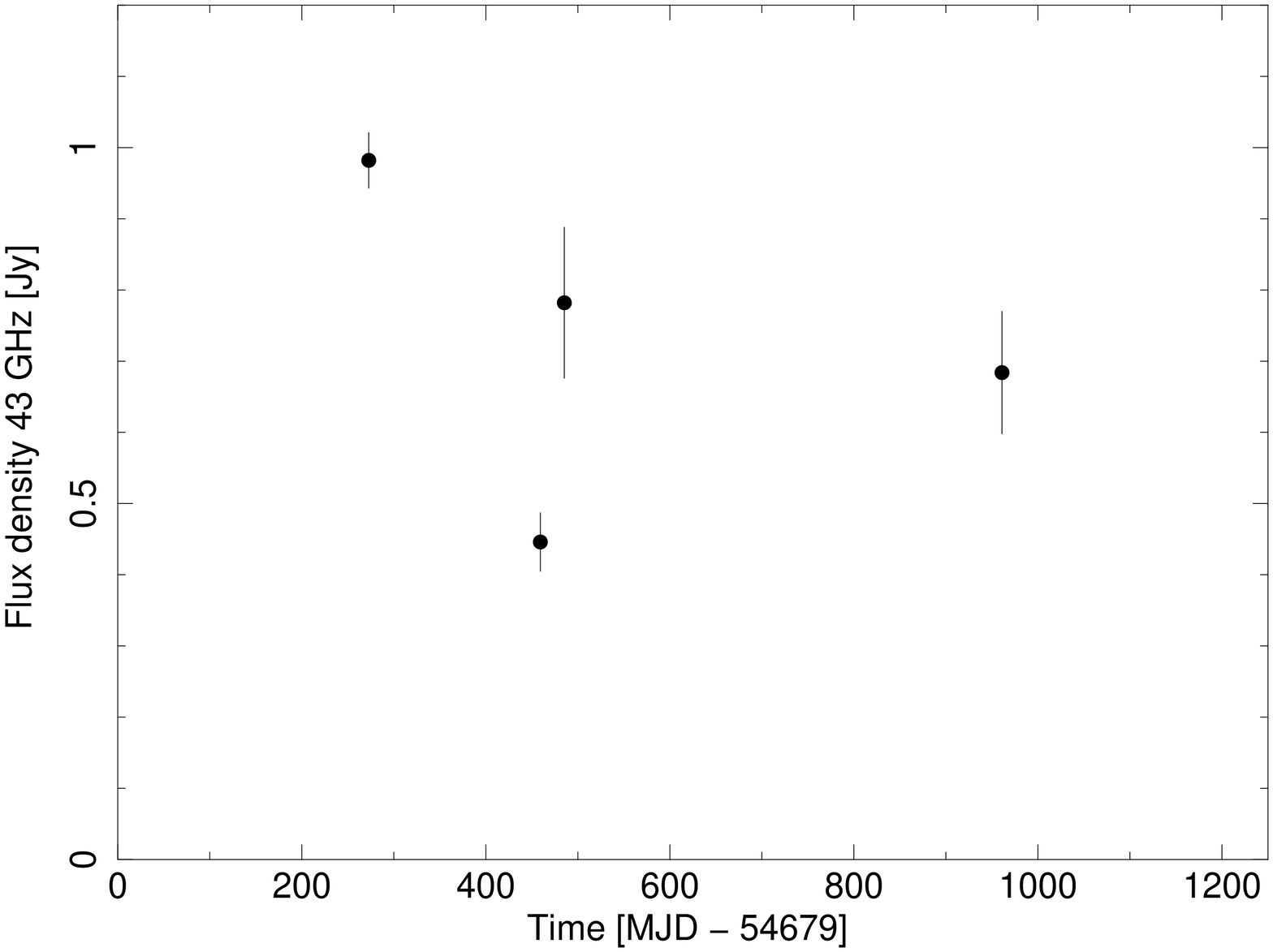}\\
\caption{Light curves at various frequencies. From top left to bottom right panel: B from {\it Swift}/UVOT [mJy]; V from {\it Swift}/UVOT [mJy]; J from INAOE [mJy]; H from INAOE [mJy]; K from INAOE [mJy]; 142 GHz from IRAM [Jy]; 86 GHz from IRAM [Jy]; 43 GHz from Effelsberg [Jy]. Time starts on 2008 August 1 00:00 UTC (MJD 54679).}
\label{Fig:CURVE2}
\end{figure*}

\begin{figure*}
\centering
\includegraphics[angle=270,scale=0.3]{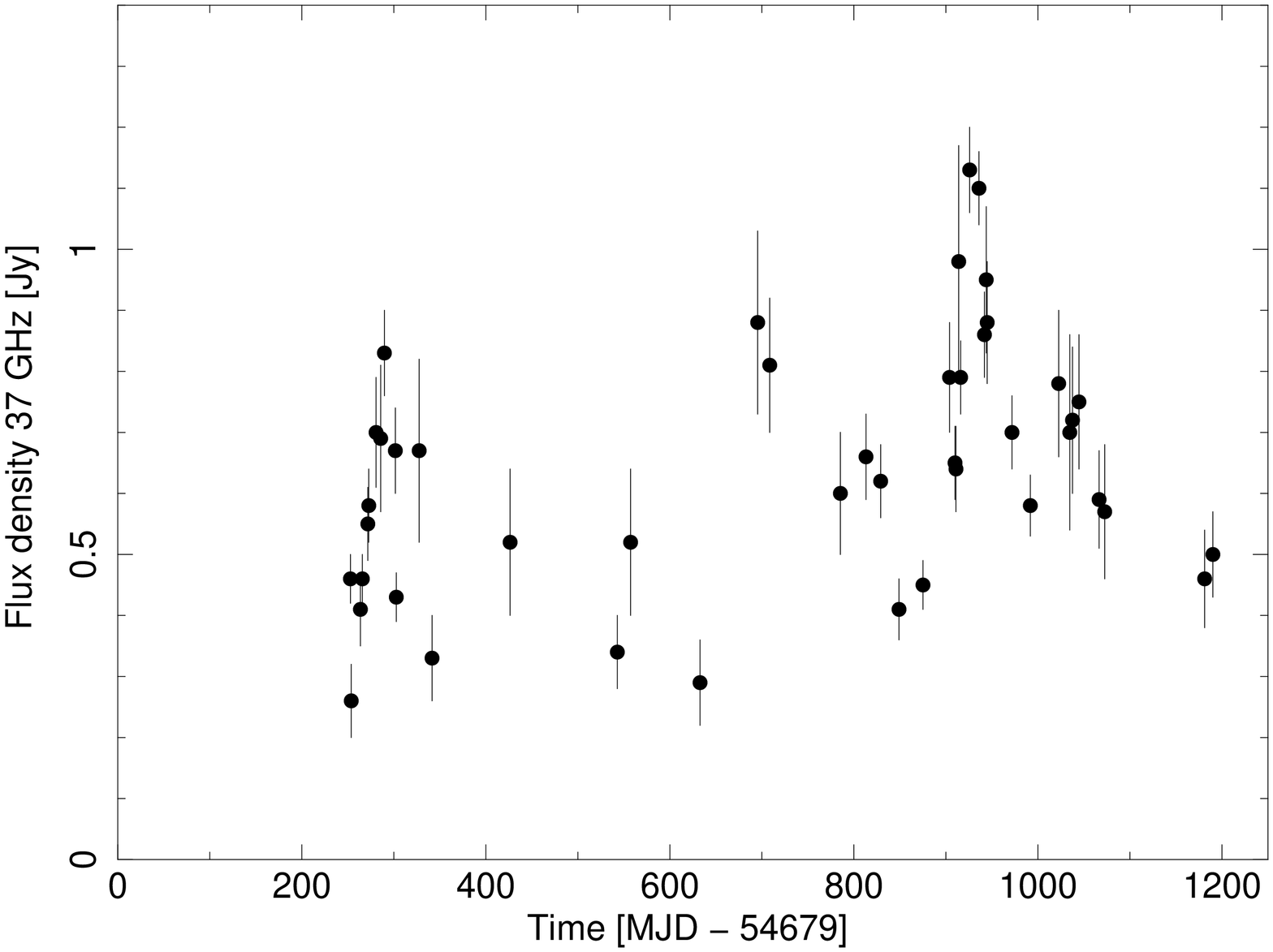}
\includegraphics[angle=270,scale=0.3]{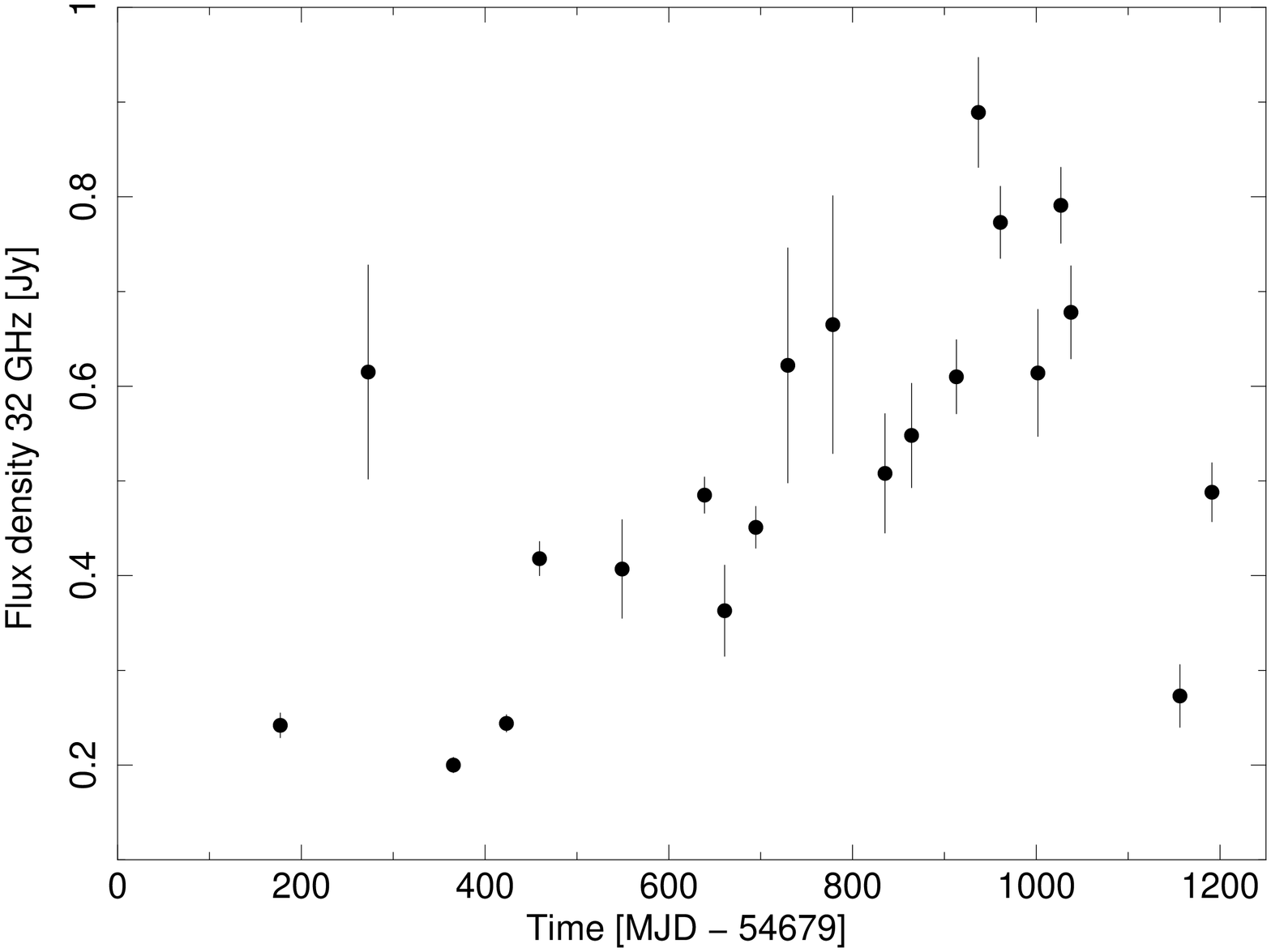}\\
\includegraphics[angle=270,scale=0.3]{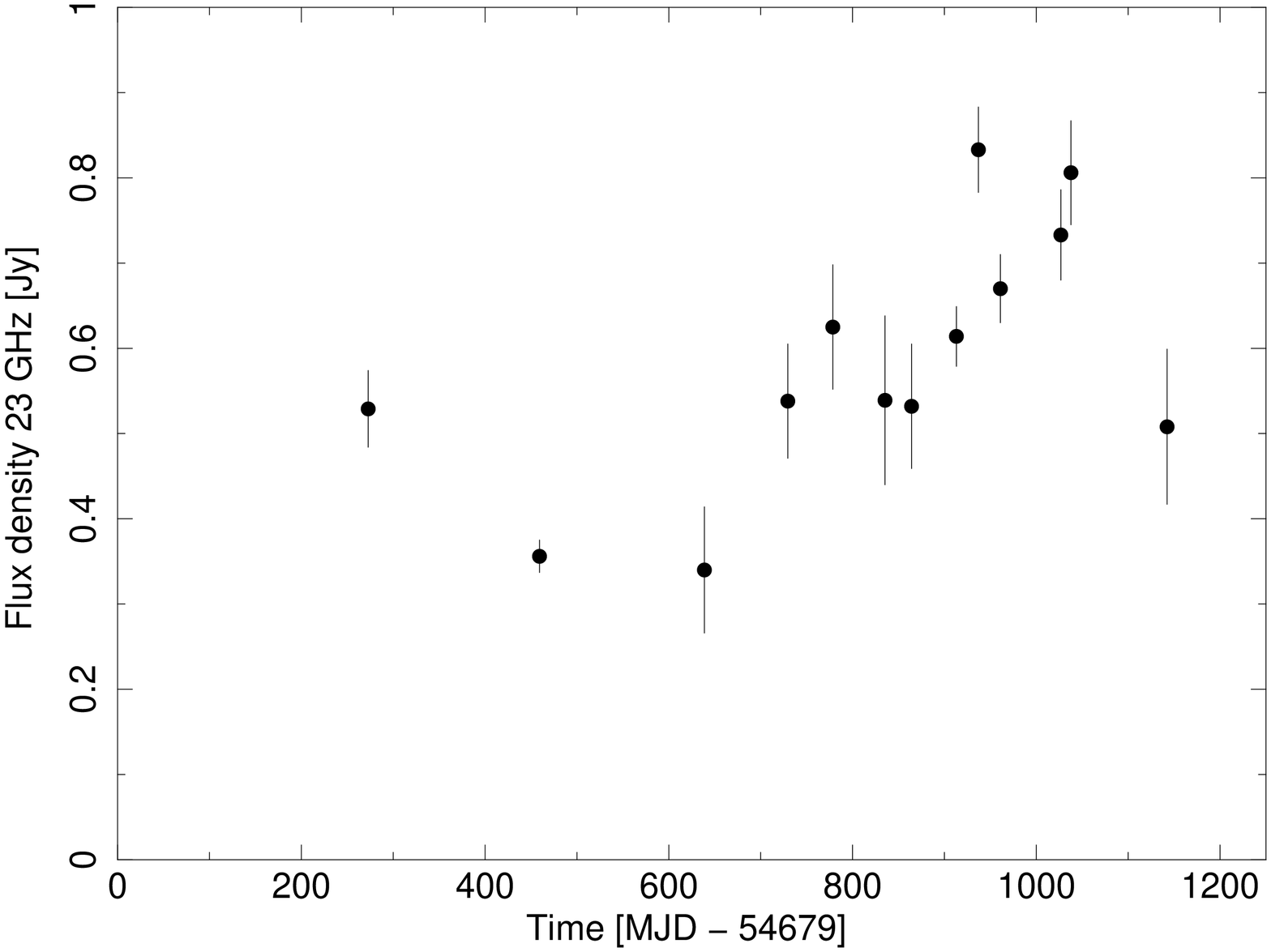}
\includegraphics[angle=270,scale=0.3]{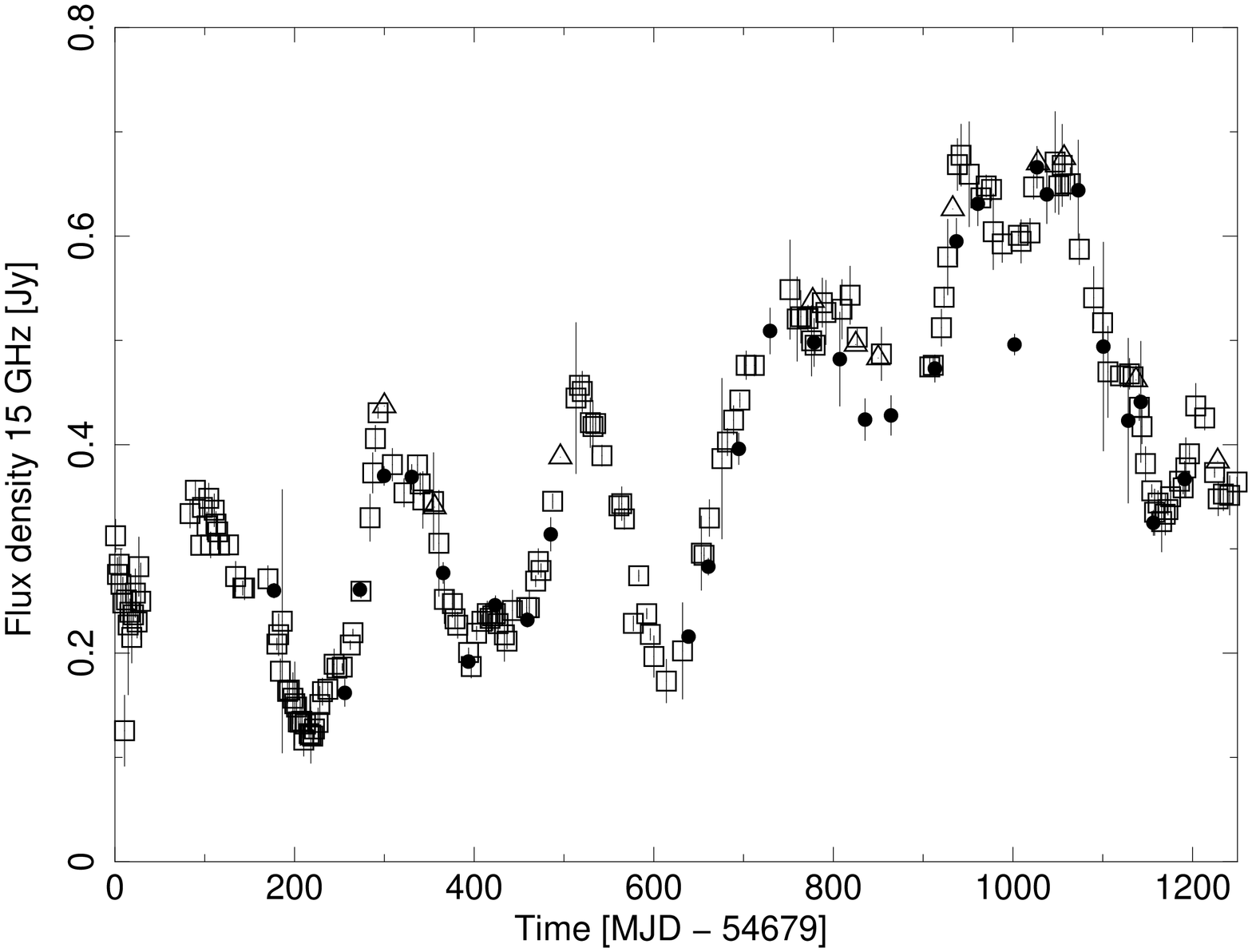}\\
\includegraphics[angle=270,scale=0.3]{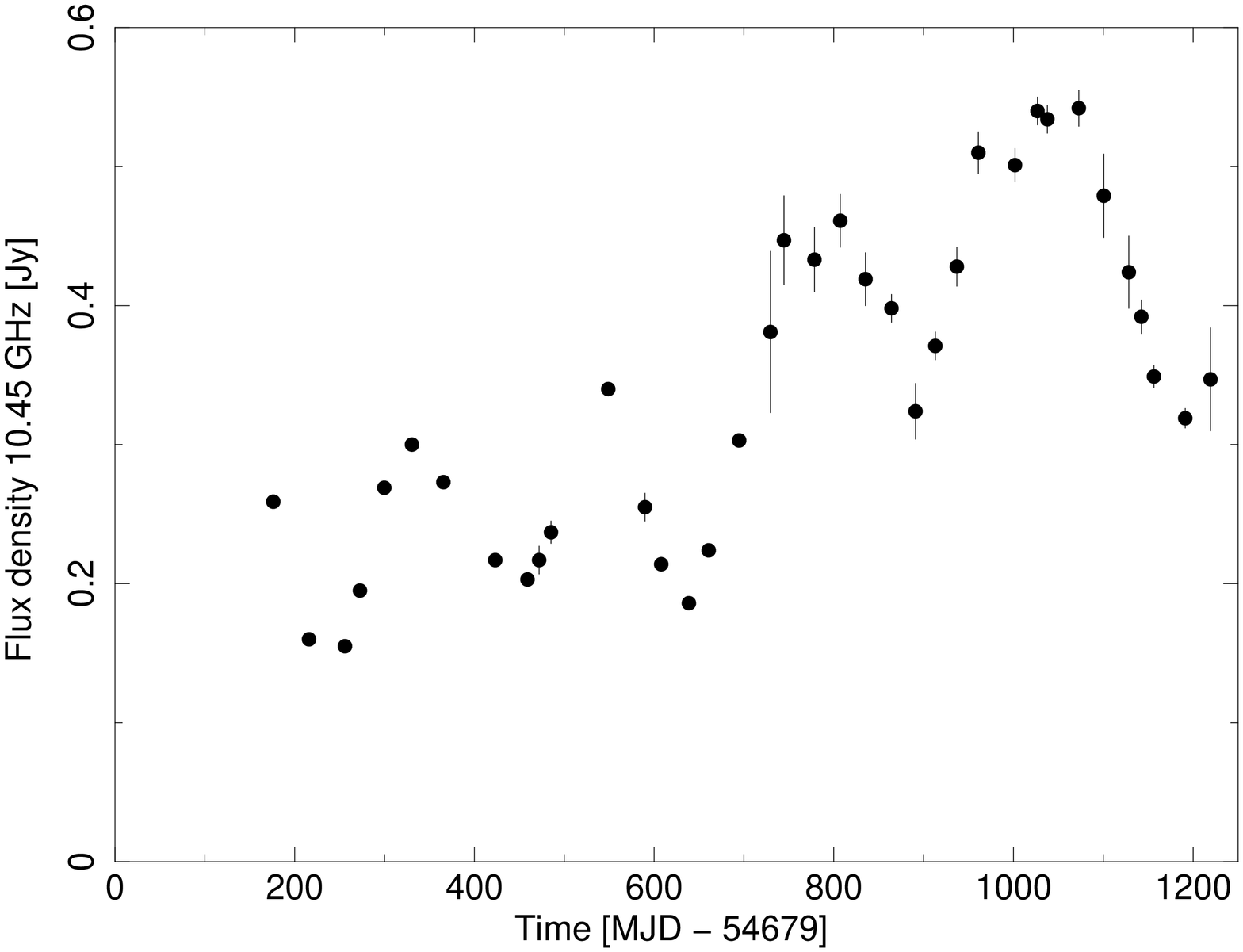}
\includegraphics[angle=270,scale=0.3]{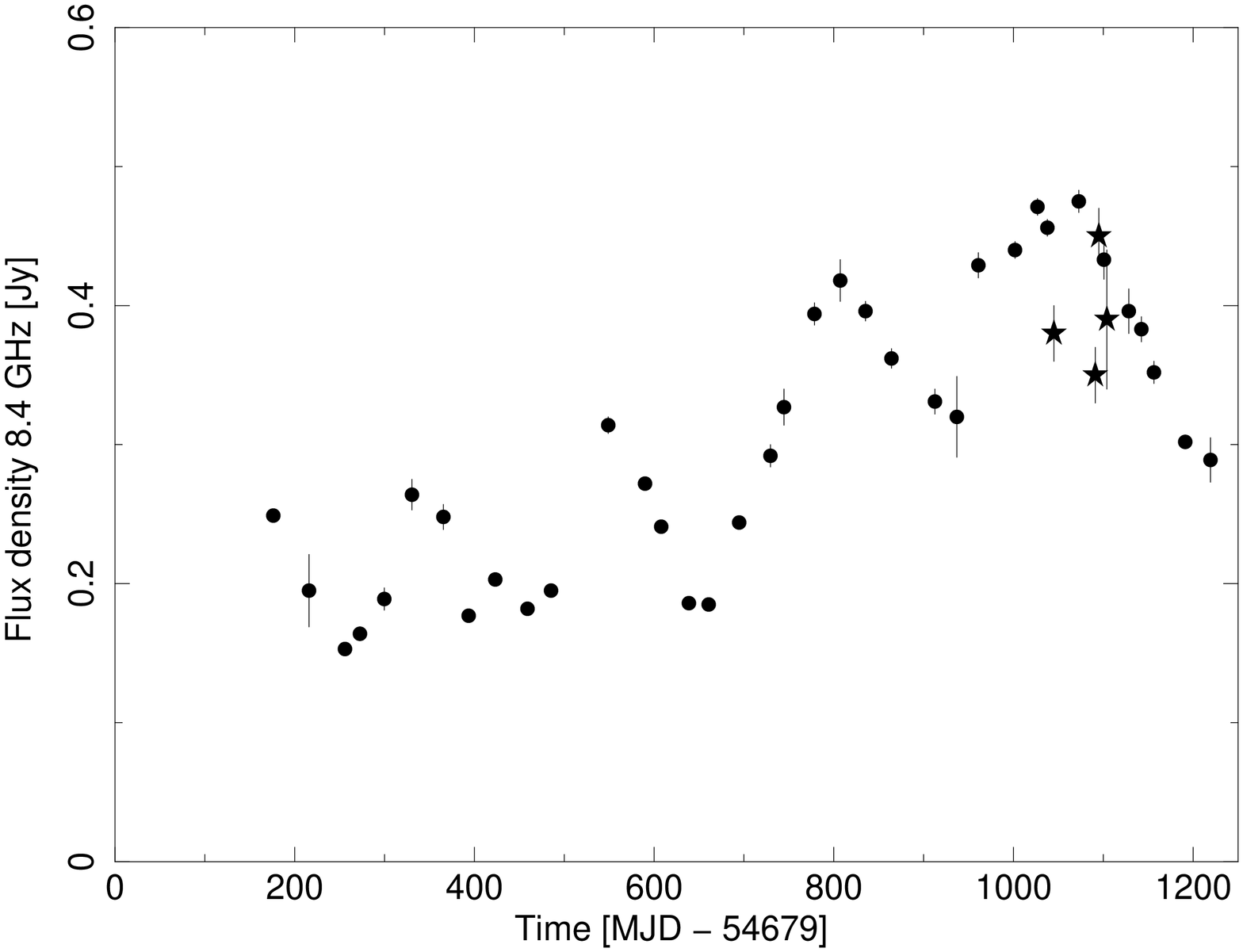}\\
\includegraphics[angle=270,scale=0.3]{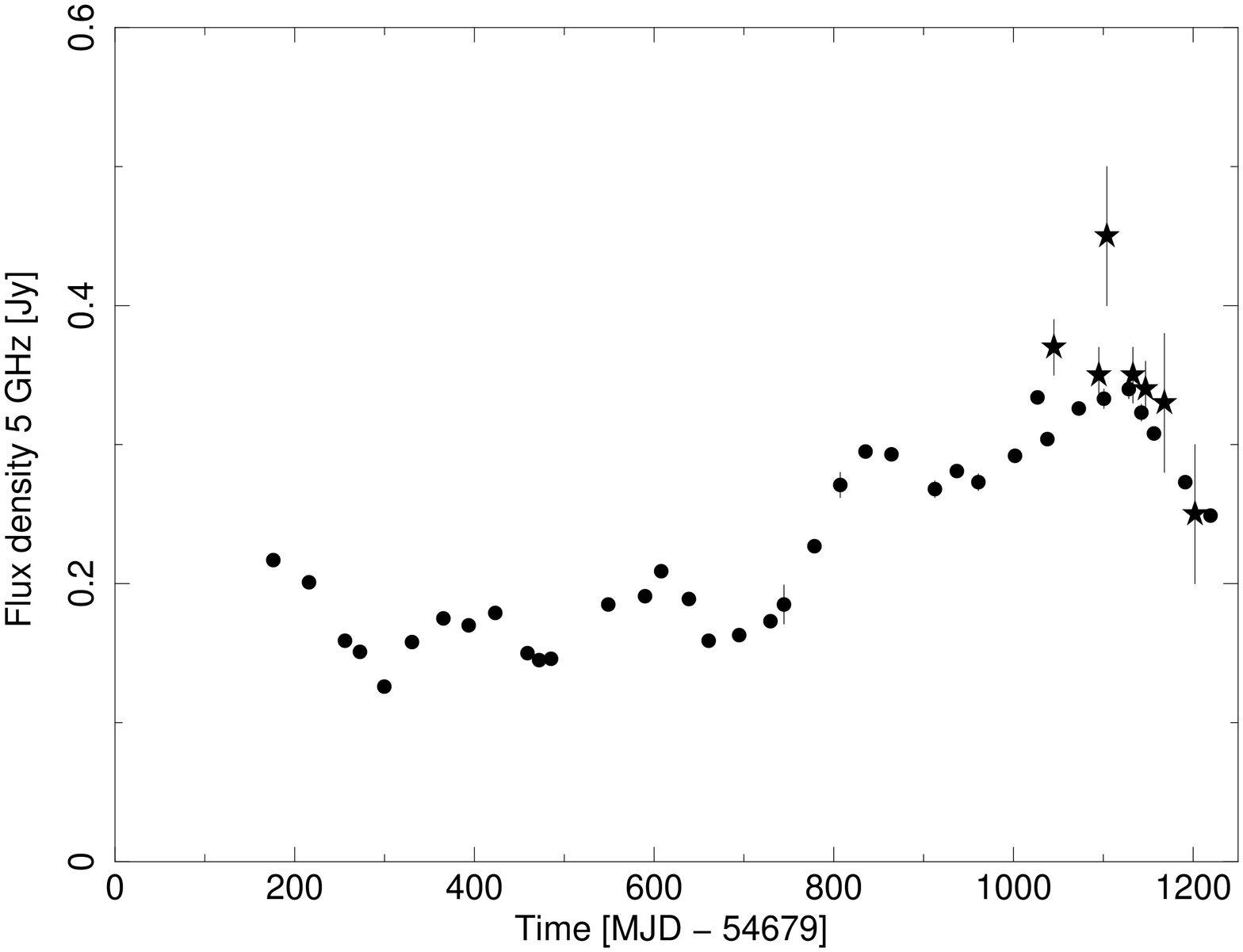}
\includegraphics[angle=270,scale=0.3]{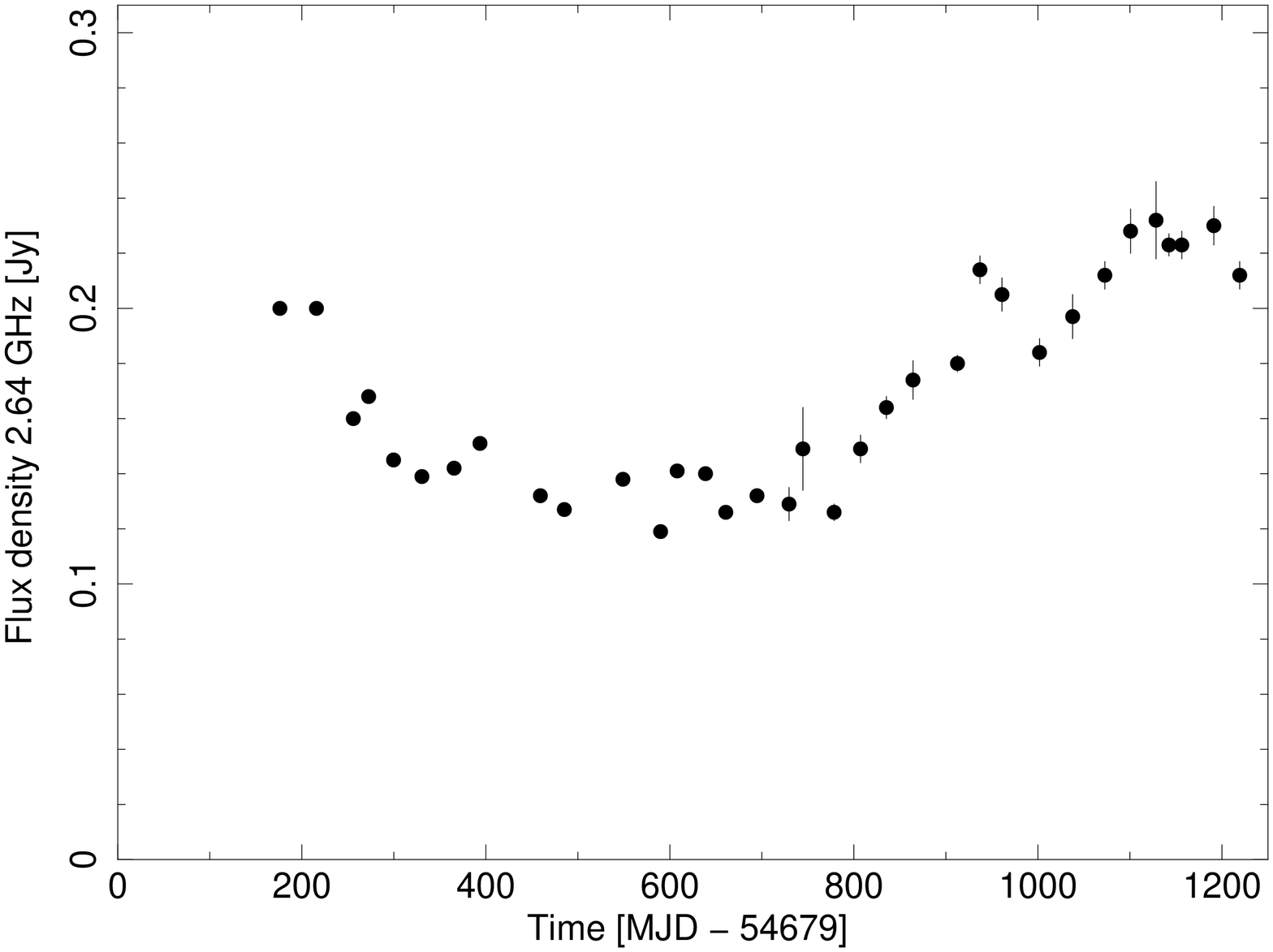}
\caption{Light curves at various frequencies. From top left to bottom right panel: 37 GHz from Mets\"ahovi [Jy]; 32 GHz from Effelsberg [Jy]; 23 GHz from Effelsberg [Jy]; 15 GHz from OVRO (open squares), Effelsberg (filled circles), and MOJAVE (open triangles) [Jy]; 10.45 GHz from Effelsberg [Jy]; 8.4 GHz from Effelsberg (filled circles) and Medicina (filled stars) [Jy]; 5.0 GHz from Effelsberg (filled circles) and Medicina (filled stars) [Jy]; 2.64 GHz  from Effelsberg [Jy]. Time starts on 2008 August 1 00:00 UTC (MJD 54679).}
\label{Fig:CURVE3}
\end{figure*}
\end{appendix}


\begin{thebibliography}{99}
\bibitem[AGILE Coll. (2011)]{AGILE} AGILE Collaboration (Lucarelli F. et al.), 2011, Astron. Telegram, 3448

\bibitem[Angelakis et al.(2008)]{ANGELAKIS} Angelakis, E., Fuhrmann, L., Marchili, N., Krichbaum, T.~P., \& Zensus, J.~A.\ 2008, Mem. Soc. Astron. Ita. 79, 1042

\bibitem[Angelakis et al. (2012)]{ANGELAKIS2} Angelakis, E., et al., 2012, Unification and physical interpretation of the radio spectra variability patterns in Fermi blazars and jet emission from NLSy1s. In: ``Fermi \& Jansky: Our evolving understanding of AGN'', St. Michaels (MD, USA), 10-12 November 2011, eConf C111110 ({\tt arXiv:1205.1961})

\bibitem[Baars et al. (1977)]{BAARS} Baars, J. W. M., Genzel, R., Paulini-Toth, I. I. K., \& Witzel, A. 1977, A\&A 61, 99

\bibitem[Barthelmy et al. (2005)]{BAT} Barthelmy, S., Barbier, L., Cummings, J., et al., 2005, Space Sci. Rev. 120, 143

\bibitem[Blandford \& Rees (1978)]{BLANDFORD2} Blandford, R. D., \& Rees, M. J., 1978, Some comments on radiation mechanisms in Lacertids. In: ``Proceedings of the Pittsburgh Conference on BL Lac Objects'', Pittsburgh (PA, USA), 24-26 April 1978, University of Pittsburgh, p. 328

\bibitem[Blandford \& K\"onigl (1979)]{BLANDFORD1} Blandford, R. D., \& K\"onigl, A., 1979, ApJ, 232, 34

\bibitem[Bonning et al. (2012)]{BONNING} Bonning, E., Urry, C. M., Bailyn, C., et al., 2012, ApJ, 756, 13

\bibitem[Breeveld et al.(2010)]{breeveld10} Breeveld, A. A., et al.,   2010, MNRAS, 406, 1687

\bibitem[Burrows et al. (2005)]{XRT} Burrows, D., Hill, J., Nousek, J., et al., 2005, Space Sci. Rev. 120, 165

\bibitem[Cardelli et al. (1989)]{CARDELLI} Cardelli, J.~A., Clayton, G.~C., \& Mathis, J.~S., 1989, ApJ 345, 245

\bibitem[Cash (1979)]{CASH} Cash, W. 1979, ApJ, 228, 939

\bibitem[Cusumano et al. (2010)]{CUSUMANO} Cusumano. G., et al., 2010, A\&A, 524, A64

\bibitem[Doi et al. (2006)]{DOI} Doi, A., et al., 2006, PASJ, 58, 829

\bibitem[Foschini et al. (2007)]{LUIGI} Foschini, L., Ghisellini, G., Tavecchio, F., et al., 2007, Swift follow-up of the gigantic TeV ourburst of PKS 2155$-$304 in 2006. In: ``The First GLAST Symposium'', ed. S. Ritz, P. Michelson \& C. Meegan, AIP Conf. Proc. (Melville, NY), 921, p. 329

\bibitem[Foschini (2010)]{outburst2} Foschini L., 2010, Astron. Telegram, 2752 

\bibitem[Foschini (2011a)]{nls1ws} Foschini, L., 2011a, Evidence of powerful relativistic jets in Narrow-Line Seyfert 1 Galaxies, in: ``Narrow-Line Seyfert 1 Galaxies and Their Place in the Universe'', ed. L. Foschini, M. Colpi, L. Gallo, D. Grupe, S. Komossa, K. Leighly \& S. Mathur, Proceedings of Science (Trieste), vol. NLS1, id 024.

\bibitem[Foschini (2011b)]{FOSCHINI1} Foschini, L., 2011b, Res. Astron. Astrophys. 11, 1266 

\bibitem[Foschini (2012)]{FOSCHINI2} Foschini, L., 2012, Powerful relativistic jets in spiral galaxies, in: ``Proceedings of the Conference on High Energy Phenomena in Relativistic Outflows III (HEPRO III)'', ed J. M. Paredes, M. Rib\'o, F. A. Aharonian \& G. E. Romero, Int. J. Mod. Phys. Conf. Series, 8, 172.

\bibitem[Foschini et al. (2011a)]{0948camp2010} Foschini, L., Ghisellini, G., Kovalev, Y.Y. et al., 2011a, MNRAS 413, 1671

\bibitem[Foschini et al. (2011b)]{0948camp2010bis} Foschini, L., Ghisellini, G., Maraschi, L. et al., 2011b, The July 2010 outburst of the NLS1 PMN J0948+0022. In: ``The Third Fermi Symposium'', Roma (Italy), 9-12 May 2011, eConf C110509, {\tt arXiv:1110.5649}

\bibitem[Foschini et al. (2011c)]{FASTVAR} Foschini, L., Ghisellini, G., Tavecchio, F., Bonnoli, G., \& Stamerra, A., 2011c, A\&A, 530, A77

\bibitem[Fuhrmann et al.(2007)]{2007AIPC..921..249F} Fuhrmann, L., Zensus, J.~A., Krichbaum, T.~P., Angelakis, E., \& Readhead, A.~C.~S., 2007, Simultaneous radio to (sub-) mm-monitoring of variability and spectral shape evolution of potential GLAST blazars, in: ``The First GLAST Symposium'', ed. S. Ritz, P. Michelson, \& C. A. Meegan, AIP Conf. Proc. (Melville, NY), vol. 921, p. 249

\bibitem[Fuhrmann et al.(2008)]{2008A&A...490.1019F} Fuhrmann, L., et al.\ 2008, A\&A 490, 1019

\bibitem[Ghisellini \& Tavecchio (2009)]{GGMODEL} Ghisellini, G., \& Tavecchio, F., 2009, MNRAS, 397, 985

\bibitem[Ghisellini et al. (2010)]{GHISELLINI} Ghisellini, G., Tavecchio, F., Foschini, L., Ghirlanda, G., Maraschi, L., Celotti, A., 2010, MNRAS, 402, 497

\bibitem[Giroletti et al. (2011)]{GIROLETTI} Giroletti, M., et al., 2011, A\&A, 528, L11

\bibitem[Goodrich (1989)]{GOODRICH} Goodrich, R. W., 1989, ApJ 342, 224

\bibitem[Hamilton \& Foschini (2012)]{HAMILTON} Hamilton, T. S., \& Foschini, L., 2012, American Astronomical Society, AAS Meeting \#220, \#335.07 

\bibitem[Heinz \& Sunyaev (2003)]{HEINZ} Heinz, S., \& Sunyaev, R. A., 2003, MNRAS, 343, L59

\bibitem[Hill et al. (2004)]{hill04} Hill, J.E., et al., 2004, Proc. SPIE, 5165, 217

\bibitem[Ikejiri et al. (2011)]{IKEJIRI} Ikejiri, Y., et al., 2011, PASJ, 63, 639

\bibitem[Kalberla et al. (2005)]{KALBERLA} Kalberla, P. M. W., et al., 2005, A\&A, 440, 775

\bibitem[Kirhakos et al. (1999)]{KIRHAKOS} Kirhakos, S., et al., 1999, ApJ, 520, 67

\bibitem[Komatsu et al. (2007)]{KOMATSU} Komatsu, E., et al., 2011, ApJS 192, 18

\bibitem[Komossa et al. (2006)]{KOMOSSA} Komossa, S., et al., 2006, AJ, 132, 531

\bibitem[Krawczynski et al. (2004)]{ORPHAN} Krawczynski, H., et al., 2004, ApJ, 601, 151

\bibitem[LAT Coll. (2009a)]{LAT} LAT Collaboration (Atwood, W. B. et al.), 2009a, ApJ 697, 1071

\bibitem[LAT Coll. (2009b)]{0948discovery} LAT Collaboration (Abdo, A. A. et al.), 2009b, ApJ 699, 976
\bibitem[LAT Coll. (2009c)]{0948camp2009} LAT Collaboration (Abdo, A. A. et al.), 2009c, ApJ 707, 727
\bibitem[LAT Coll. (2009d)]{nls1class} LAT Collaboration (Abdo, A. A. et al.), 2009d, ApJ 707, L142
\bibitem[LAT Coll. (2010a)]{0948disco2} LAT Collaboration (Foschini, L. et al.), 2010a, Fermi/LAT discovery of gamma-ray emission from a relativistic jet in the narrow-line Seyfert 1 quasar PMN J0948+0022, in: ``Accretion and Ejection in AGN: a Global View'', ed. L. Maraschi, G. Ghisellini, R. Della Ceca, \& F. Tavecchio, ASP Conf. Series (San Francisco, CA), vol. 427, p. 243

\bibitem[LAT Coll. (2010b)]{outburst} LAT Collaboration (Donato D. et al.), 2010b, Astron. Telegram, 2733

\bibitem[LAT Coll. (2010c)]{1FGL} LAT Collaboration (Abdo A. A. et al.), 2010c, ApJS, 188, 405


\bibitem[LAT Coll. (2010d)]{3C279} LAT Collaboration (Abdo A. A. et al.), 2010d, Nature, 463, 919

\bibitem[LAT Coll. (2010e)]{VAR} LAT Collaboration (Abdo A. A. et al.), 2010e, ApJ, 722, 520

\bibitem[LAT Coll. (2012)]{2FGL} LAT Collaboration (Nolan P. L. et al.), 2012, ApJS, 199, 31



\bibitem[Lister \& Homan (2005)]{POLARIZZAZIONE} Lister M.L. \& Homan D.C., 2005, AJ, 130, 1389

\bibitem[Lister et al. (2009)]{LISTER} Lister, M. L., et al., 2009, AJ, 137, 3718

\bibitem[Lister et al. (2011)]{LISTER2} Lister, M. L., Aller, M., Aller, H., et al., 2011, ApJ, 742, 27

\bibitem[Liu et al. (2010)]{LIU} Liu, H., Wang, J., Mao, Y. \& Wei, J., 2010, ApJ, 715, L113

\bibitem[Marscher (2012)]{MARSCHER1} Marscher, A. P., 2012, Multi-waveband Variations of Blazars during Gamma-ray Outbursts. In: ``The Third Fermi Symposium'', Roma (Italy), 9-12 May 2011, eConf C110509, ({\tt arXiv:1201.5402}).

\bibitem[Marscher et al. (2012)]{MARSCHER2} Marscher, A. P., Jorstad, S. G., Agudo, I., MacDonald, N. R., \& Scott, T. L., 2012, Relation between Events in the Millimeter-wave Core and Gamma-ray Outbursts in Blazar Jets. In: ``Fermi \& Jansky: Our evolving understanding of AGN'', St. Michaels (MD, USA), 10-12 November 2011, eConf C111110 ({\tt arXiv:1204.6707}).

\bibitem[Massaro et al. (2004)]{MASSARO} Massaro, F., Perri, M., Giommi, P., Nesci, R., 2004, A\&A, 413, 489

\bibitem[Mattox et al. (1996)]{MATTOX} Mattox, J. R., Bertsch, D. L., Chiang, J., et al., 1996, ApJ 461, 396

\bibitem[Osterbrock \& Pogge (1985)]{OSTPOGGE} Osterbrock, D. E. \& Pogge, R. W., 1985, ApJ 297, 166

\bibitem[Poole et al.(2008)]{poole08} Poole, T. S., et al., 2008, MNRAS, 383, 627

\bibitem[Richards et al. (2011)]{RICHARDS} Richards, J. L. et al. 2011, ApJS 194, 29

\bibitem[Roming et al. (2005)]{UVOT} Roming, P., Kennedy, T., Mason, K., et al., 2005, Space Sci. Rev. 120, 95

\bibitem[Scargle (1981)]{SCARGLE} Scargle, J. D., 1981, ApJS, 45, 1

\bibitem[Singal (2009)]{SINGAL} Singal, A. K., 2009, ApJ, 703, L109

\bibitem[Ter\"asranta et al. (1998)]{METSAHOVI} Ter\"asranta, H., Tornikoski, M., Mujunen, A. et al., 1998, A\&AS, 132, 305

\bibitem[Tramacere et al. (2007)]{TRAMACERE} Tramacere, A., Massaro, F., Cavaliere, A., 2007, A\&A, 466, 521

\bibitem[Tsang \& Kirk (2007)]{TSANG} Tsang, O., \& Kirk, J. G., 2007, A\&A, 463, 145

\bibitem[Ulvestadt et al. (1995)]{ULVESTADT} Ulvestadt, J. S., Antonucci, R. R. J., Goodrich, R. W., 1995, AJ, 109, 81

\bibitem[VERITAS Coll. (2011)]{VERITAS} VERITAS Collaboration (Acciari V. A. et al.), 2011, ApJ, 738, 25

\bibitem[Wagner \& Witzel (1995)]{WW} Wagner, S., \& Witzel, A., 1995, ARA\&A, 33, 163

\bibitem[Williams et al. (2002)]{RIKWILLIAMS} Williams, R. J., Pogge, R. W., Mathur, S., 2002, AJ, 124, 3042 

\bibitem[Yuan et al. (2008)]{YUAN} Yuan, W., Zhou, H.-Y., Komossa, S., et al., 2008, ApJ, 685, 801

\bibitem[Zhou et al. (2003)]{ZHOU} Zhou, H.-Y., Wang, T.-G., Dong, X.-B., Zhou, Y.-Y., \& Li, C., 2003, ApJ 584, 147
\end{thebibliography}
\end{document}